\documentclass[aps,prd,preprint,nofootinbib,showpacs,showkeys,preprintnumbers,amsmath,amssymb,11pt]{revtex4}

\usepackage{graphicx}
\usepackage{dcolumn}
\usepackage{bm}
\usepackage{bbm}
\usepackage{multirow}

\allowdisplaybreaks

\begin{document}


\title{Neutrino Oscillation in Dense Matter\vspace{3mm}}

\author{\bf Shu Luo}
\email{luoshu@xmu.edu.cn}
\affiliation{Department of Astronomy, Xiamen University, Xiamen, Fujian 361005, China \vspace{9mm}}

\begin{abstract}
\vspace{3mm}
As the increasing of neutrino energy or matter density, the neutrino oscillation in matter may undergo ``vacuum-dominated'', ``resonance'' and ``matter-dominated'' three different stages successively. Neutrinos endure very different matter effects, and therefore present very different oscillation behaviors in these three different cases. In this paper, we focus on the less discussed matter-dominated case (i.e., $|A^{}_{\rm CC}| \gg |\Delta m^{2}_{31}|$), study the effective neutrino mass and mixing parameters as well as neutrino oscillation probabilities in dense matter using the perturbation theory. We find that as the matter parameter $|A^{}_{\rm CC}|$ growing larger, the effective mixing matrix in matter $\tilde{V}$ evolves approaching a fixed $3 \times 3$ constant real matrix which is free of CP violation and can be described using only one simple mixing angle $\tilde{\theta}$ which is independent of $A^{}_{\rm CC}$. As for the neutrino oscillation behavior, $\nu^{}_{e}$ decoupled in the matter-dominated case due to its intense charged-current interaction with electrons while a two-flavor oscillation are still presented between $\nu^{}_{\mu}$ and $\nu^{}_{\tau}$. Numerical analysis are carried on to help understanding the salient features of neutrino oscillation in matter as well as testing the validity of those concise approximate formulas we obtained. At the end of this paper, we make a very bold comparison of the oscillation behaviors between neutrinos passing through the Earth and passing through a typical white dwarf to give some embryo thoughts on under what circumstances these studies will be applied and put forward the interesting idea of possible ``neutrino lensing'' effect.
\vspace{3mm}
\end{abstract}



\maketitle

\section{Introduction}

When neutrinos pass through a medium, the interactions with the particles in the background give rise to modifications of the properties of neutrinos as well as the oscillation behaviors. This is well known as the matter effect which have been playing important roles in understanding various neutrino oscillation data. In the standard three neutrinos framework, the effective Hamiltonian $\tilde{\cal{H}}$ in the flavor basis responsible for the propagation of neutrinos in matter, differs from the Hamiltonian in vacuum $\cal{H}$,
\begin{eqnarray}
\tilde{\cal H} & = & {\cal H} + {\cal H}' \; = \; \frac{1}{2 E} \left [ V \left ( \begin{matrix} m^{2}_{1} & & \cr & m^{2}_{2} & \cr & & m^{2}_{3} \cr \end{matrix} \right ) V^{\dagger}_{} + \left ( \begin{matrix} A^{}_{\rm CC} + A^{}_{\rm NC} & & \cr & A^{}_{\rm NC} & \cr & & A^{}_{\rm NC} \cr \end{matrix} \right ) \right ] \; ,
\end{eqnarray}
where $\cal{H}' $ describes the forward coherent scattering of neutrinos with the constituents of the medium (i.e., electrons, protons and neutrons) via the weak charged-current (CC) and neutral-current (NC) interactions \cite{Wolfenstein:1977ue, Mikheev:1986gs, Mikheev:1986wj, Kuo:1989qe}. Here $A^{}_{\rm CC} = 2 E V^{}_{\rm CC}$, $A^{}_{\rm NC} = 2 E V^{}_{\rm NC}$ (with $V^{}_{\rm CC} = \sqrt{2} G^{}_{\rm F} N^{}_{e}$ and $\displaystyle V^{}_{\rm NC} = - \frac{\sqrt{2}}{2} G^{}_{\rm F} N^{}_{n}$ being the effective matter potentials) are parameters of the same unit as the mass-squared difference $\Delta m^{2}_{ji}$ that measure the strength of the matter effect, and $V$ is just the $3 \times 3$ unitary Pontecorvo-Maki-Nakagawa-Sakata (PMNS) leptonic mixing matrix \cite{Maki:1962mu, Pontecorvo:1967fh} which is conventionally parametrized  in terms of three mixing angles ${ \theta^{}_{12}, \theta^{}_{13}, \theta^{}_{23} }$ and one Dirac CP-violating phase $\delta$ as \cite{Tanabashi:2018oca}
\begin{eqnarray}
V \; = \; \left ( \begin{matrix} c^{}_{12} c^{}_{13} & s^{}_{12} c^{}_{13} & s^{}_{13} e^{-i\delta}_{} \cr -s^{}_{12} c^{}_{23} - c^{}_{12} s^{}_{23} s^{}_{13} e^{i\delta}_{} & c^{}_{12} c^{}_{23} - s^{}_{12} s^{}_{23} s^{}_{13} e^{i\delta}_{} & s^{}_{23} c^{}_{13} \cr s^{}_{12} s^{}_{23} - c^{}_{12} c^{}_{23} s^{}_{13} e^{i\delta}_{} & - c^{}_{12} s^{}_{23} - s^{}_{12} c^{}_{23} s^{}_{13} e^{i\delta}_{} & c^{}_{23} c^{}_{13} \end{matrix} \right ) \; ,
\end{eqnarray}
where $c^{}_{ij} \equiv \cos\theta^{}_{ij}$ and $s^{}_{ij} \equiv \sin\theta^{}_{ij}$ (for $ij = 12, 13, 23$) have been introduced. Throughout this paper we do not consider the possible Majorana phases, simply because they are irrelevant to neutrino oscillations in both vacuum and matter.
For anti-neutrino oscillation in matter, one may simply replace $V$ by $V^{*}_{}$ and $A^{}_{\rm CC}$ by $-A^{}_{\rm CC}$ in the effective Hamiltonian (i.e., $A^{}_{\rm CC}$ is negative in the case of anti-neutrino oscillation). 

The intriguing matter effect is a result of the interplay between the vacuum Hamiltonian ${\cal H}$ and the matter term ${\cal H}'$. Note that, the diagonal term $\displaystyle \frac{1}{2 E} \left ( m^{2}_{1} + A^{}_{\rm NC} \right ) \cdot \mathbbm{1}$ in Eq. (1) develops just a common phase for all three flavors, and does not affect the neutrino oscillation behaviors. Therefore it is the interplay among the two mass-squared differences $\Delta m^{2}_{21}$, $\Delta m^{2}_{31}$, the mixing parameters in $V$ (which are all parameters in the vacuum Hamiltonian ${\cal H}$ and have been well determined from varieties of neutrino oscillation experiments \cite{Tanabashi:2018oca, Esteban:2018azc}) and the matter term $A^{}_{\rm CC}$ (which will vary with the matter density and the energy of neutrino) that give rise to varied neutrino oscillation behaviors.

According to the relative magnitude of $\Delta m^{2}_{21}$, $|\Delta m^{2}_{31}|$ and $|A^{}_{\rm CC}|$, the various possible values of $A^{}_{\rm CC}$ can be laid in three main different regions: {\bf the vacuum-dominated region} (i.e., $|A^{}_{\rm CC}| \ll \Delta m^{2}_{21}, |\Delta m^{2}_{31}|$), {\bf the resonance region} (i.e., $A^{}_{\rm CC} \sim \Delta m^{2}_{21}, \Delta m^{2}_{31}$), and {\bf the matter-dominated region} (i.e., $|A^{}_{\rm CC}| \gg \Delta m^{2}_{21}, |\Delta m^{2}_{31}|$). Among various studies of the matter effect, neutrino oscillation behaviors in the resonance region attracted the most attention (see e.g., \cite{Zaglauer:1988gz, Cervera:2000kp, Freund:2001pn, Xu:2015kma, Li:2016pzm, Denton:2018hal}). The oscillation probabilities may get dramatic corrections owing to the resonances at around $A^{}_{\rm CC} \sim \Delta m^{2}_{21}$ (solar resonance) or $A^{}_{\rm CC} \sim \Delta m^{2}_{31}$ (atmospheric resonance) \footnote{The more accurate resonance conditions in the two-flavor or three-flavor neutrino oscillation picture are discussed in e.g., Ref. \cite{Akhmedov:2004ny}.} which are crucially important for the studies of atmospheric neutrinos, accelerator neutrino beams passing through the Earth or the spectrum of solar neutrinos. Also there have been discussions concerning the vacuum-dominated case \cite{Barger:1980tf, Blennow:2003xw, Parke:2016joa, Xing:2016ymg, Denton:2016wmg}, which could be helpful for various long- or medium-baseline neutrino oscillation experiments with neutrino beam energy $E$ below the  solar resonance. In this region, the neutrino oscillation probabilities as well as the leptonic CP violation receive predictable small corrections from the matter effect.

Recently interests have been shown in exploring the less discussed matter-dominated case \cite{Xing:2018lob, Huang:2018ufu, Wang:2019yfp, Xing:2019owb}, where the matter term ${\cal H}'$ dominates over the vacuum Hamiltonian ${\cal H}$, or more specifically, $|A^{}_{\rm CC}| \gg |\Delta m^{2}_{31}|$. Such studies are applicable in the case of neutrinos having extremely high energy or going through extremely dense object. Further to these works, we explore in this paper the effective neutrino mass and mixing parameters as well as the neutrino oscillation probabilities in dense matter using the perturbation theory. We find that as the matter parameter $|A^{}_{\rm CC}|$ growing larger, the effective mixing matrix in matter $\tilde{V}$ evolves approaching a fixed $3 \times 3$ constant real matrix which is free of CP violation and can be described using simply one mixing angle $\tilde{\theta}$ which is independent of the matter parameter $A^{}_{\rm CC}$. As for the neutrino oscillation behavior, $\nu^{}_{e}$ decoupled in the matter-dominated case due to its intense charged-current interaction with electrons in the medium while a two-flavor oscillation can still present between $\nu^{}_{\mu}$ and $\nu^{}_{\tau}$. As far as the six neutrino oscillation parameters in vacuum are well determined and the condition $|A^{}_{\rm CC}| \gg |\Delta m^{2}_{31}|$ is satisfied, the neutrino oscillation probabilities in dense matter can be well predicted regardless if the matter density varies along the path. 

We plan to organize the remaining parts of this paper as following. In section II we aim to reveal the features of the effective neutrino masses and mixing matrix in matter under the condition $|A^{}_{\rm CC} / \Delta m^{2}_{31}| \rightarrow \infty$ with the help of the perturbation theory. Base on the results of the series expansions, a set of pretty concise approximate formulas of neutrino oscillation probabilities in the matter-dominated region are derived in section III. Numerical analysis are carried on in both sections to help understanding the salient features of neutrino oscillation in matter as $|A^{}_{\rm CC}|$ changes from zero to infinity as well as testing the validity of those concise formulas. Finally, in section IV we make a very bold comparison of the oscillation behaviors between neutrinos passing through the Earth and passing through a typical white dwarf so as to answer the question under what circumstances these studies will be applied and put forward the interesting idea of possible ``neutrino lensing'' effect.

\section{Fixed points of the effective neutrino mass and mixing parameters in the matter-dominated case}

As already mentioned above, in the standard three neutrinos framework, the effective Hamiltonian $\tilde{\cal{H}}$ in the flavor basis responsible for the propagation of neutrinos in matter can be written as
\begin{eqnarray}
\tilde{\cal{H}} & = & \frac{1}{2 E} \left [ \left ( m^{2}_{1} +  A^{}_{\rm NC} \right) \cdot \mathbbm{1} + V \left ( \begin{matrix} ~ 0 ~ & & \cr & \Delta m^{2}_{21} & \cr & & \Delta m^{2}_{31} \cr \end{matrix} \right ) V^{\dagger}_{} + \left ( \begin{matrix} A^{}_{\rm CC} & & \cr & ~ 0 ~ & \cr & & ~ 0 ~ \cr \end{matrix} \right ) \right ] \; \nonumber\\[1mm]
& \equiv & \frac{1}{2 E} \tilde{V} \left ( \begin{matrix} \tilde{m}^{2}_{1} & & \cr & \tilde{m}^{2}_{2} & \cr & & \tilde{m}^{2}_{3} \cr \end{matrix} \right ) \tilde{V}^{\dagger}_{} \; ,
\end{eqnarray}
where the effective neutrino masses $\tilde{m}^{}_{i}$ (for $i = 1, 2, 3$) and flavor mixing matrix $\tilde{V}$ in mater have been defined. Given a constant matter profile, the exact analytical relations between \{$\tilde{V}$, $\tilde{m}^{}_{i}$\} and \{$V$, $m^{}_{i}$\} have been established in many works using different approaches \cite{Barger:1980tf, Zaglauer:1988gz, Xing:2000gg, Xing:2001yg, Xing:2003ez, Chiu:2010da, Zhou:2016luk, Chiu:2017ckv}. And the neutrino oscillation probabilities in matter can be written in the same way as those in vacuum by simply replacing $V^{}_{\alpha i}$ and $\Delta m^{2}_{ji}$ with the corresponding effective parameters $\tilde{V}^{}_{\alpha i}$ and $\Delta \tilde{m}^{2}_{ji}$. As for any realistic profile of the matter density, it is also possible to numerically calculate the neutrino oscillation probabilities by solving the evolution equations of neutrino flavor states. 

However, in the matter-dominated region we are concerning, some useful and more transparent analytical approximations could be obtained by regarding both $\Delta m^{2}_{21} / |A^{}_{\rm CC}|$ and $|\Delta m^{2}_{31} / A^{}_{\rm CC}|$ as small parameters and performing the diagonalization of $\tilde{\cal H}$ using the perturbation theory. 
In comparison with the method adopted in previous works \cite{Xing:2018lob, Huang:2018ufu, Wang:2019yfp, Xing:2019owb} that applying further simplification on the exact formulas, the series expansion method can automatically achieve approximate formulas with any required accuracy. Moreover, the values those effective parameters in matter would approach in the limit $|A^{}_{\rm CC}| \rightarrow \infty$ are straightforwardly given in the zeroth order expansion. Also note that, different from previous works on series expansions \cite{Cervera:2000kp, Freund:2001pn, Akhmedov:2004ny, Xu:2015kma, Minakata:2015gra, Denton:2016wmg, Li:2016pzm, Denton:2018hal, Parke:2018brr, Denton:2018fex} which usually regard known constant such as $\alpha \equiv \Delta m^{2}_{21} / \Delta m^{2}_{31}$ or $\sin\theta^{}_{13}$ as small expansion parameters, the two expansion parameters $\Delta m^{2}_{21} / |A^{}_{\rm CC}|$ and $|\Delta m^{2}_{31} / A^{}_{\rm CC}|$ we employed in this paper vary with the matter parameter $A^{}_{\rm CC}$, i.e., vary with neutrino energy $E$ as well as the matter density $\rho$. As a result, this kind of series expansion relates only to the matter-dominated case, and the accuracies of those approximate formulas given in this paper depend also on the magnitude of $A^{}_{\rm CC}$. We will have a detailed discussion on this problem later at the end of Sec. III.
The details of the diagonalization of the effective Hamiltonian $\tilde{\cal{H}}$ are given in Appendix A, where the approximate expressions of three eigenvalues of $\tilde{\cal{H}}$, the effective mixing matrix and the neutrino oscillation probabilities in matter up to the first order of both $\Delta m^{2}_{21} / |A^{}_{\rm CC}|$ and $|\Delta m^{2}_{31} / A^{}_{\rm CC}|$ are also presented. 

As the increase of $|A^{}_{\rm CC}|$,  terms proportional to $1 / A^{}_{\rm CC}$ are all approaching zero, and as one can clearly seen from Eqs. (A12) and (A13), three eigenvalues of $\tilde{\cal H}$ are approaching a set of fixed values
\begin{eqnarray}
\tilde{\lambda}^{fixed}_{1} & = & \frac{1}{2 E} \left ( m^{2}_{1} + A^{}_{\rm NC} + A^{}_{\rm CC} + \Omega^{}_{11} \right ) \; , \nonumber\\[1mm]
\tilde{\lambda}^{fixed}_{2} & = & \frac{1}{2 E} \left ( m^{2}_{1} + A^{}_{\rm NC} + \Omega^{}_{22} \cos^2\tilde{\theta} +\Omega^{}_{33} \sin^2\tilde{\theta} - | \Omega^{}_{23} | \sin2\tilde{\theta} \right ) \; , \nonumber\\[1mm]
\tilde{\lambda}^{fixed}_{3} & = & \frac{1}{2 E} \left ( m^{2}_{1} + A^{}_{\rm NC} + \Omega^{}_{33} \cos^2\tilde{\theta} +\Omega^{}_{22} \sin^2\tilde{\theta} + | \Omega^{}_{23} | \sin2\tilde{\theta} \right ) \; ,
\end{eqnarray}
where the Hermitian matrix $\Omega$ is defined as
\begin{eqnarray}
\Omega & \equiv & V \left ( \begin{matrix} ~ 0 ~ & & \cr & \Delta m^{2}_{21} & \cr & & \Delta m^{2}_{31} \cr \end{matrix} \right ) V^{\dagger}_{} \;,
\end{eqnarray}
and the complete expressions of its nine elements $\Omega^{}_{ij}$ can be found in Eq. (A8). Apparently, in this matter-dominated case, $\tilde{\lambda}^{fixed}_{2}$ and $\tilde{\lambda}^{fixed}_{3}$ are nearly degenerate and both of them have strong hierarchies with $\tilde{\lambda}^{fixed}_{1}$.
In the same time the effective mixing matrix in matter $\tilde{V}$ evolves towards a fixed $3 \times 3$ real matrix
\begin{eqnarray}
\tilde{V}^{fixed}_{} & = & \left ( \begin{matrix} ~ 1 ~ & ~ 0 ~ & ~ 0 ~ \cr ~ 0 ~ & \cos \tilde{\theta} & \sin \tilde{\theta} \cr ~ 0 ~ & - \sin \tilde{\theta} & \cos \tilde{\theta} \cr \end{matrix} \right )  \; ,
\end{eqnarray}
which has the two-flavor-mixing structure and can be parametrized using just one mixing angle $\tilde{\theta}$ defined by
{\small
\begin{eqnarray}
\tan 2 \tilde{\theta} & = & \frac{2 \left | \Delta m^{2}_{21} V^{}_{\mu 2} V^{*}_{\tau 2}  + \Delta m^{2}_{31} V^{}_{\mu 3} V^{*}_{\tau 3} \right |}{\Delta m^{2}_{21} \left ( | V^{}_{\tau 2} |^{2}_{} - | V^{}_{\mu 2} |^{2}_{} \right ) + \Delta m^{2}_{31} \left ( | V^{}_{\tau 3} |^{2}_{} - | V^{}_{\mu 3} |^{2}_{} \right )} \nonumber\\[3mm]
& = & \frac{\left | \Delta m^{2}_{21} \left ( \sin2\theta^{}_{23} ( s^{2}_{12} s^{2}_{13} - c^{2}_{12} ) - \cos2\theta^{}_{23} \sin2\theta^{}_{12} s^{}_{13} \cos\delta + i \sin2\theta^{}_{12} s^{}_{13} \sin\delta \right ) + \Delta m^{2}_{31}  \sin2\theta^{}_{23} c^{2}_{13} \right |}{\Delta m^{2}_{21} \left ( \cos2\theta^{}_{23} ( s^{2}_{12} s^{2}_{13} - c^{2}_{12} ) + \sin2\theta^{}_{23} \sin2\theta^{}_{12} s^{}_{13} \cos\delta \right ) + \Delta m^{2}_{31} \cos2\theta^{}_{23} c^{2}_{13} } \; . \nonumber\\
\end{eqnarray}}Considering the strong hierarchy of $\Delta m^{2}_{21} \ll |\Delta m^{2}_{31}|$ and the smallness of $s^{}_{13}$, one can immediately obtain from above equation that the mixing angle $\tilde{\theta} \approx \theta^{}_{23}$ \footnote{The approximate relation is confirmed by our numerical analysis. In the normal mass ordering case we have $\tilde{\theta} = 49.83^{\circ}$ ($\tan\tilde{\theta} = 1.196$) while in the inverted mass ordering case we have $\tilde{\theta} = 49.70^{\circ}$ ($\tan\tilde{\theta} = 1.179$), both are very close to the input value of $\theta^{}_{23} = 49.7^{\circ}$ ($\tan\theta^{}_{23} = 1.179$) as one can see later in Figs. 3 and 4.}. One may also find that the mixing angle $\tilde{\theta}$ defined in Eq. (6) is actually an indicator of the $\mu$-$\tau$ symmetry breaking in the Dirac neutrino mass matrix \footnote{For a recent review on the $\mu$-$\tau$ symmetry, see e.g., Ref. \cite{Xing:2015fdg}}. If the neutrino mass matrix in vacuum $M \equiv V {\rm diag}\{ m^{2}_{1}, m^{2}_{2}, m^{2}_{3} \} V^{\dagger}_{}$ possess the exact $\mu$-$\tau$ symmetry, we then have $\tilde{\theta} = \pi/4$. 

The fixed points in the limit $|A^{}_{\rm CC}| \rightarrow \infty$ has been noticed in Refs. \cite{Xing:2018lob, Wang:2019yfp, Xing:2019owb, Parke:2018brr}, in which the evolution behaviors of not only nine elements of the effective mixing matrix $|\tilde{V}^{}_{\alpha i}|$ but also those mass and mixing parameters are illustrated. It's worth to go a step further drawing a full picture of the evolution behaviors of three effective neutrino masses and the effective mixing matrix in the matter-dominated case. In the limit $|A^{}_{\rm CC} / \Delta m^{2}_{31} | \rightarrow \infty$, one of the eigenstates of $\tilde{\lambda}^{}_{1}$ is decoupled due to the large potential of $A^{}_{\rm CC}$ and $A^{}_{\rm NC}$ while the other two eigenvalues are nearly degenerate ($\tilde{\lambda}^{}_{2} \simeq \tilde{\lambda}^{}_{3}$) for they are both dominated by the large neutral-current potential term $A^{}_{\rm NC}$. Correspondingly, the $3 \times 3$ effective mixing matrix $\tilde{V}$ in matter presents a nearly two-flavor-mixing structure. It means $\tilde{V}$ asymptotically conserves intrinsic CP and can be well described by just one mixing angles $\tilde{\theta}$, which can be approximately expressed as $\tilde{\theta} \approx \theta^{}_{23}$.

\begin{table}
\caption{The best-fit values of six neutrino oscillation parameters from a global fit of current experiment data \cite{Esteban:2018azc}}
\begin{tabular}{ccccccc}
\hline\noalign{\smallskip}
& ~~~ $\theta^{}_{12}$ ~~~ & ~~~ $\theta^{}_{13}$ ~~~ & ~~~ $\theta^{}_{23}$ ~~~ & ~~~ $\delta$\; ~~~ & ~ $\Delta m^{2}_{21} ~ [10^{-5} {\rm eV}^{2}]$ ~ & ~ $\Delta m^{2}_{31} ~ [10^{-3} {\rm eV}^{2}]$ ~ \\
\noalign{\smallskip}\hline\noalign{\smallskip}
Normal Mass Ordering ~~ & ~ $33.82^\circ$ ~ & ~ $8.61^\circ$ ~ & ~ $49.7^\circ$ ~ & ~ $217^\circ$ ~ & ~ $7.39$ ~ & ~~$2.451$ \\
\noalign{\smallskip}
Inverted Mass Ordering ~~ & ~ $33.82^\circ$ ~ & ~ $8.65^\circ$ ~ & ~ $49.7^\circ$ ~ & ~ $280^\circ$ ~ & ~ $7.39$ ~ & $-2.512$ \\
\noalign{\smallskip}\hline
\end{tabular}
\end{table}

\begin{figure}
\begin{center}
\vspace{0cm}
\includegraphics[width=.7\textwidth]{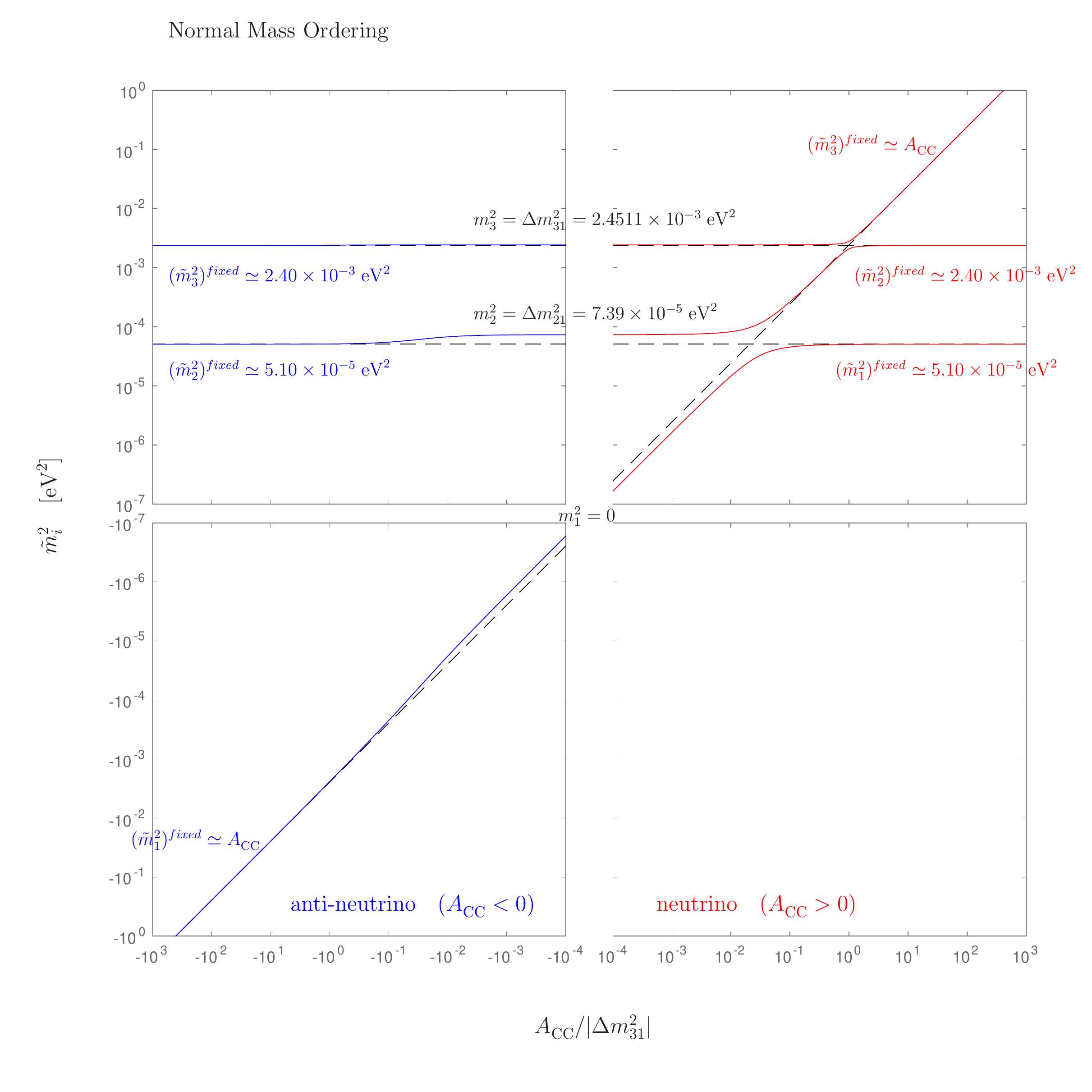}
\vspace{0cm}
\caption{The evolution of three squared effective neutrino masses $\tilde{m}^{2}_{i}$ (for $i = 1, 2, 3$) in matter with respect to the dimensionless ratio $A^{}_{\rm CC} / |\Delta m^{2}_{31}|$ in the normal mass ordering case for both neutrinos (with $A^{}_{\rm CC} > 0$, red curves in the right half panel) and anti-neutrinos (with $A^{}_{\rm CC} < 0$, blue curves in the left half panel) , where the best-fit values of the mass-squared differences and the mixing parameters in Table. I have been input. Note that, the common terms $m^{2}_{1} + A^{}_{\rm NC}$ are omitted from all three $\tilde{m}^{2}_{i}$ for the sake of simplicity, while the relation $\Delta \tilde{m}^{2}_{ji} = \tilde{m}^{2}_{j} - \tilde{m}^{2}_{i}$ still holds. Both the input values of $\tilde{m}^{2}_{i}$ in vacuum and the fixed points in the limit $|A^{}_{\rm CC}| \gg |\Delta m^{2}_{31}|$ are given on the plots.}
\end{center}
\end{figure}

\begin{figure}
\begin{center}
\vspace{0cm}
\includegraphics[width=.7\textwidth]{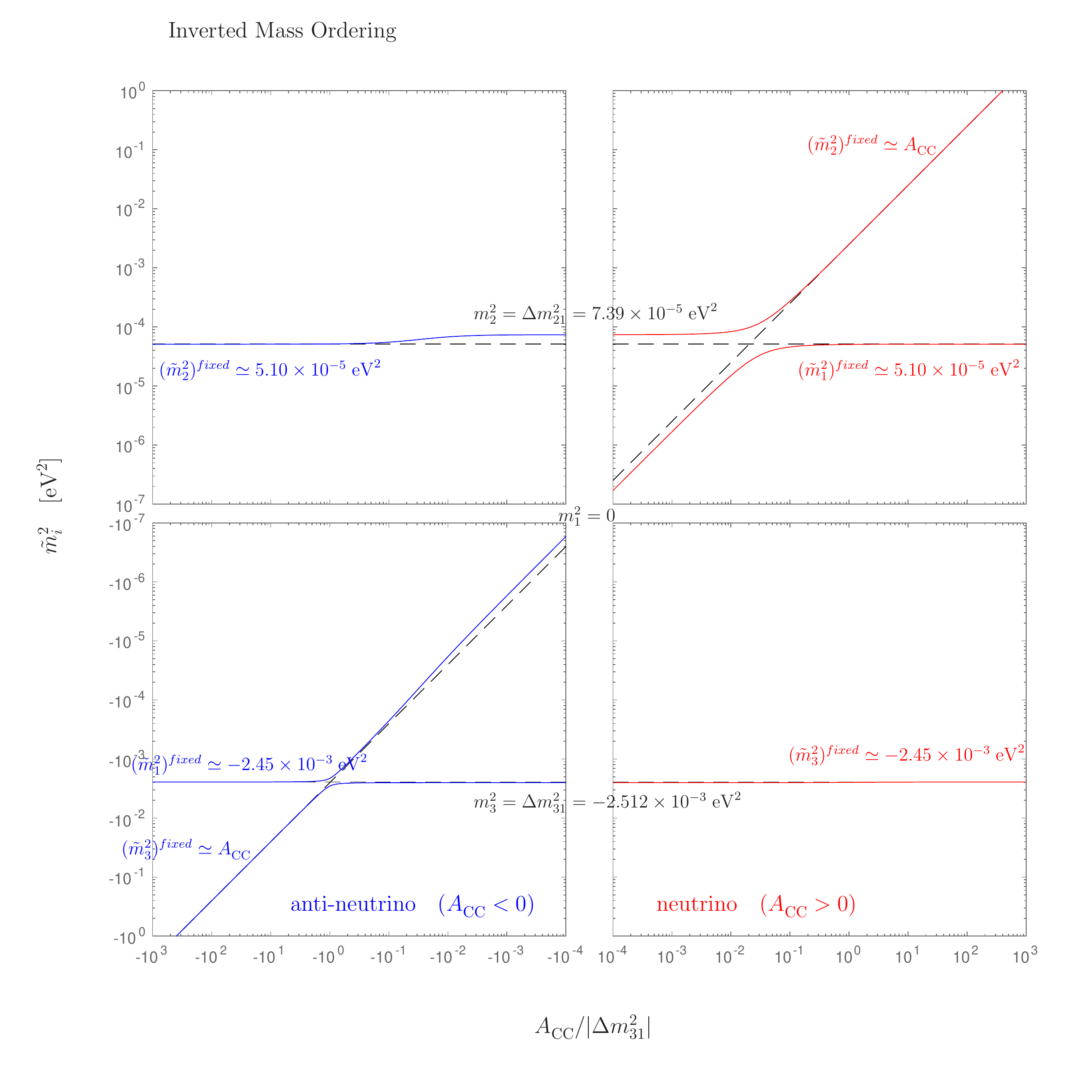}
\vspace{0cm}
\caption{The evolution of three squared effective neutrino masses $\tilde{m}^{2}_{i}$ (for $i = 1, 2, 3$) in matter with respect to the dimensionless ratio $A^{}_{\rm CC} / |\Delta m^{2}_{31}|$ in the inverted mass ordering case for both neutrinos (with $A^{}_{\rm CC} > 0$, red curves in the right half panel) and anti-neutrinos (with $A^{}_{\rm CC} < 0$, blue curves in the left half panel) , where the best-fit values of the mass-squared differences and the mixing parameters in Table. I have been input. Note that, the common terms $m^{2}_{1} + A^{}_{\rm NC}$ are omitted from all three $\tilde{m}^{2}_{i}$ for the sake of simplicity, while the relation $\Delta \tilde{m}^{2}_{ji} = \tilde{m}^{2}_{j} - \tilde{m}^{2}_{i}$ still holds. Both the input values of $\tilde{m}^{2}_{i}$ in vacuum and the fixed points in the limit $|A^{}_{\rm CC}| \gg |\Delta m^{2}_{31}|$ are given on the plots.}
\end{center}
\end{figure}

To see the features of fixed points as well as the evolution of $\tilde{m}^{}_{i}$ and $\tilde{V}$ more transparently, we illustrate in Figs. 1 and 2 the evolution of three squared effective neutrino masses in matter $\tilde{m}^{2}_{i} = 2 E \tilde{\lambda}^{}_{i}$ (for $i = 1, 2, 3$), in Figs. 3 and 4 the evolution of the modulus of nine elements of thr effective mixing matrix in matter $|\tilde{V}^{}_{\alpha i}|$ (for $\alpha = e, \mu, \tau$ and $i = 1, 2, 3$) and in Fig. 5 the effective Jarlskog $\tilde{\cal J} \equiv {\rm Im} \left ( V^{}_{\alpha i} V^{}_{\beta j} V^{*}_{\alpha j} V^{*}_{\beta i} \right ) \sum^{}_{\gamma, k} \epsilon^{}_{\alpha \beta \gamma} \epsilon^{}_{ijk}$ (for $\alpha, \beta, \gamma = e, \mu, \tau$ and $i, j, k = 1, 2, 3$) \footnote{The effective Jarlskog $\tilde{\cal J}$ stands for the CP violation in the effective mixing matrix $\tilde{V}$, for more discussions on the properties of the Jarlskog in matter, see e.g. Refs. \cite{Naumov:1991ju, Toshev:1991ku, Harrison:1999df, Yokomakura:2000sv, Xing:2000ik, Parke:2000hu, Xing:2016ymg, Denton:2019yiw, Wang:2019dal}.} \cite{Jarlskog:1985ht, Wu:1985ea} with the increasing of the dimensionless ratio $A^{}_{\rm CC} / |\Delta m^{2}_{31}|$ in both the normal and the inverted mass ordering cases. The best-fit values of the neutrino oscillation parameters from Ref. \cite{Esteban:2018azc} as summarized in Table I has been adopted as the inputs in vacuum ($A^{}_{\rm CC} = 0$) in our numerical calculations. One can clearly see that the evolution behaviors of these effective parameters in matter in the region $|A^{}_{\rm CC} / \Delta m^{2}_{31}| \gg 1$ are all in good agreement with the predictions of Eqs. (4)-(7).

\begin{table}
\caption{The resulting eigenvalues $\tilde{\lambda}^{}_{i} = \tilde{m}^{2}_{i} / 2 E$ (for $i = 1, 2, 3$) of the effective Hamiltonian $\tilde{\cal H}$ and the corresponding effective mixing matrix in matter $\tilde{V}$ in the limit $|A^{}_{\rm CC}| \rightarrow \infty$ for both neutrinos and anti-neutrinos with different mass orderings. Where $\tilde{\lambda}^{fixed}_{i}$ (for $i = 1, 2, 3$) are defined in Eq. (4) and $\tilde{\theta}$ can be calculated using Eq. (7). Note that the three eigenvalues $\tilde{\lambda}^{}_{i}$ are ordered in such a way that in all four scenarios the same correct order $\{ \tilde{\lambda}^{}_{1}, \tilde{\lambda}^{}_{2}, \tilde{\lambda}^{}_{3} \} = \{ m^{2}_{1}, m^{2}_{2}, m^{2}_{3} \} / 2E$ can be obtained in the limit $A^{}_{\rm CC} = 0$ through continuous evolution as $|A^{}_{\rm CC}|$ decreasing as one can see in Figs. 1 and 2.}
\begin{tabular}{c|cc|cc}
\hline
&&&&\\[-5.5mm]
& \multicolumn{2}{c|}{~ Normal Mass Ordering ~ ($\Delta m^{2}_{31} > 0$) ~} & \multicolumn{2}{c}{~ Inverted Mass Ordering ~ ($\Delta m^{2}_{31} < 0$) ~} \\[1mm]
& \begin{tabular}{c} ~ neutrinos ~ \\[-1mm] ~ ($A^{}_{\rm CC} > 0$) ~ \\[1mm] \end{tabular} &\begin{tabular}{c} ~ anti-neutrinos ~ \\[-1mm] ~ ($A^{}_{\rm CC} < 0$) ~ \\[1mm] \end{tabular} & \begin{tabular}{c} ~ neutrinos ~ \\[-1mm] ~ ($A^{}_{\rm CC} > 0$) ~ \\[1mm] \end{tabular} & \begin{tabular}{c} ~ anti-neutrinos ~ \\[-1mm] ~ ($A^{}_{\rm CC} < 0$) ~ \\[1mm] \end{tabular} \\
\hline
\begin{tabular}{c} ~ resonances ~ \\[-1mm] ~ induced ~ \\[1mm] \end{tabular} & \begin{tabular}{c} ~ solar ~ \\[-1mm] ~ ($A ^{}_{\rm CC} \sim \Delta m^{2}_{21}$) ~ \\[1mm] ~ atmospheric ~ \\[-1mm] ~ ($A ^{}_{\rm CC} \sim \Delta m^{2}_{31}$) ~ \\[1mm] \end{tabular} & ~ None ~ & \begin{tabular}{c} ~ solar ~ \\[-1mm] ~ ($A ^{}_{\rm CC} \sim \Delta m^{2}_{21}$) ~ \\[1mm] \end{tabular} & \begin{tabular}{c} ~ atmospheric ~ \\[-1mm] ~ ($A ^{}_{\rm CC} \sim \Delta m^{2}_{31}$) ~ \\[1mm] \end{tabular} \\
\hline
&&&&\\[-5mm]
\begin{tabular}{c} ~ eigenvalues ~ \\[-1mm] ~ of $\tilde{\cal H}$ ~ \\[1mm] \end{tabular} & $\left ( \begin{matrix} ~ \tilde{\lambda}^{fixed}_{2} ~ \cr ~ \tilde{\lambda}^{fixed}_{3} ~ \cr ~ \tilde{\lambda}^{fixed}_{1} ~ \end{matrix} \right )$ & $\left ( \begin{matrix} ~ \tilde{\lambda}^{fixed}_{1} ~ \cr ~ \tilde{\lambda}^{fixed}_{2} ~ \cr ~ \tilde{\lambda}^{fixed}_{3} ~ \end{matrix} \right )$ & $\left ( \begin{matrix} ~ \tilde{\lambda}^{fixed}_{2} ~ \cr ~ \tilde{\lambda}^{fixed}_{1} ~ \cr ~ \tilde{\lambda}^{fixed}_{3} ~ \end{matrix} \right )$ & $\left ( \begin{matrix} ~ \tilde{\lambda}^{fixed}_{3} ~ \cr ~ \tilde{\lambda}^{fixed}_{2} ~ \cr ~ \tilde{\lambda}^{fixed}_{1} ~ \end{matrix} \right )$ \\
&&&&\\[-4mm]
~ $\tilde{V}^{fixed}_{}$ ~ & ~ $\left ( \begin{matrix} 0 & 0 & 1 \cr \cos\tilde{\theta} & \sin\tilde{\theta} & 0 \cr -\sin\tilde{\theta} & \cos\tilde{\theta} & 0 \cr \end{matrix} \right )$ ~ & ~ $\left ( \begin{matrix} 1 & 0 & 0 \cr 0 & \cos\tilde{\theta} & \sin\tilde{\theta} \cr 0 & -\sin\tilde{\theta} & \cos\tilde{\theta} \cr \end{matrix} \right )$ ~ & ~ $\left ( \begin{matrix} 0 & 1 & 0 \cr \cos\tilde{\theta} & 0 & \sin\tilde{\theta} \cr -\sin\tilde{\theta} & 0 & \cos\tilde{\theta} \cr \end{matrix} \right )$ ~ & ~ $\left ( \begin{matrix} 0 & 0 & 1 \cr \sin\tilde{\theta} & \cos\tilde{\theta} & 0 \cr \cos\tilde{\theta} & -\sin\tilde{\theta} & 0 \cr \end{matrix} \right )$ ~ \\[8.5mm]
\hline
\end{tabular}
\end{table}

\begin{figure}
\begin{center}
\vspace{0cm}
\includegraphics[width=\textwidth]{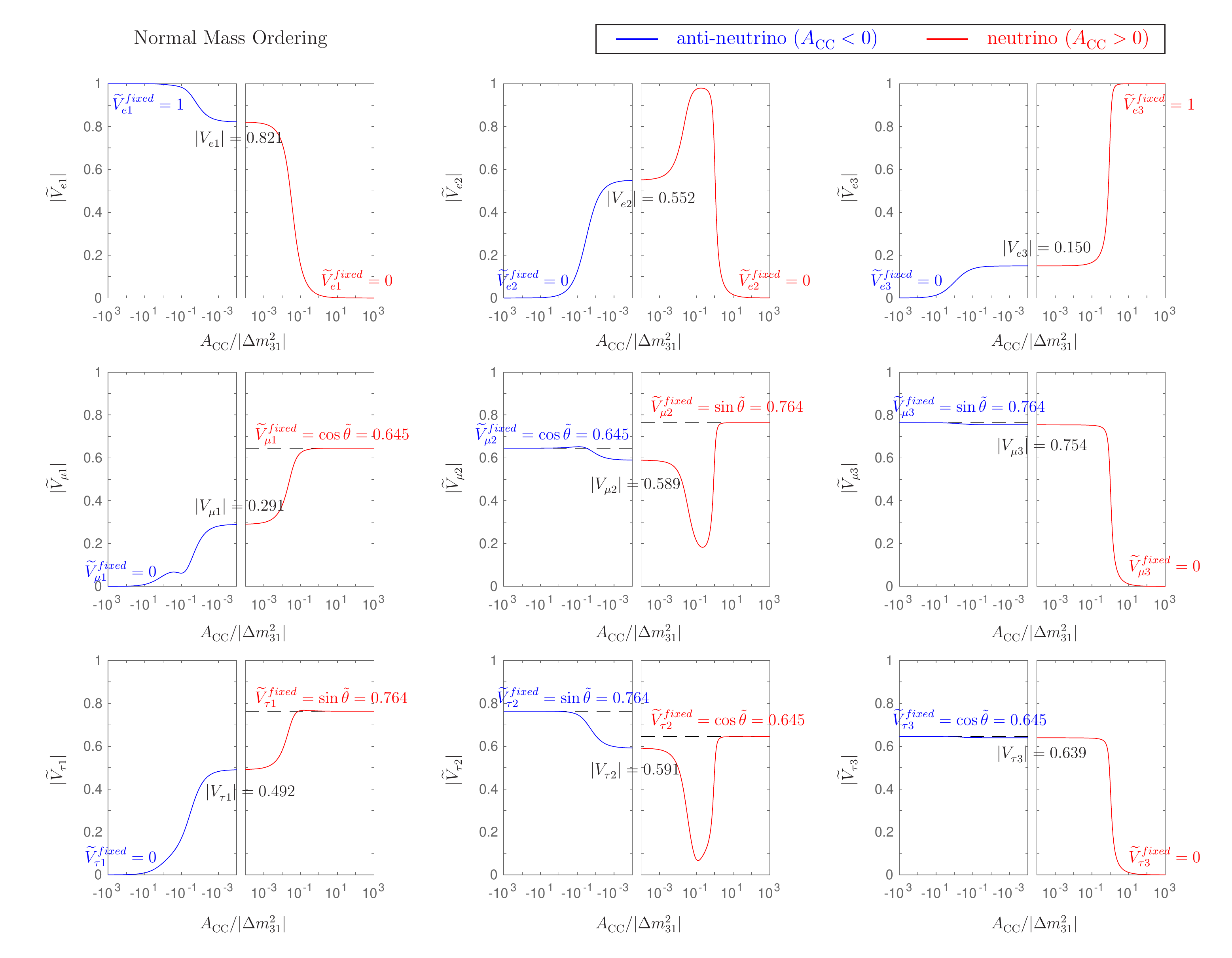}
\vspace{0cm}
\caption{The evolution of the absolute value of nine elements of the effective mixing matrix in matter $|\tilde{V}^{}_{\alpha i}|$ (for $\alpha = e, \mu, \tau$ and $i = 1, 2, 3$) with respect to the dimensionless ratio $A^{}_{\rm CC} / |\Delta m^{2}_{31}|$ in the normal mass ordering case for both neutrinos (with $A^{}_{\rm CC} > 0$, red curves in each right half panel) and anti-neutrinos (with $A^{}_{\rm CC} < 0$, blue curves in each left half panel) , where the best-fit values of the mass-squared differences and the mixing parameters in Table. I have been input. Both the input values in vacuum and the fixed points in the limit $|A^{}_{\rm CC}| \gg |\Delta m^{2}_{31}|$ are given on the plots.} 
\end{center}
\end{figure}

\begin{figure}
\begin{center}
\vspace{0cm}
\includegraphics[width=\textwidth]{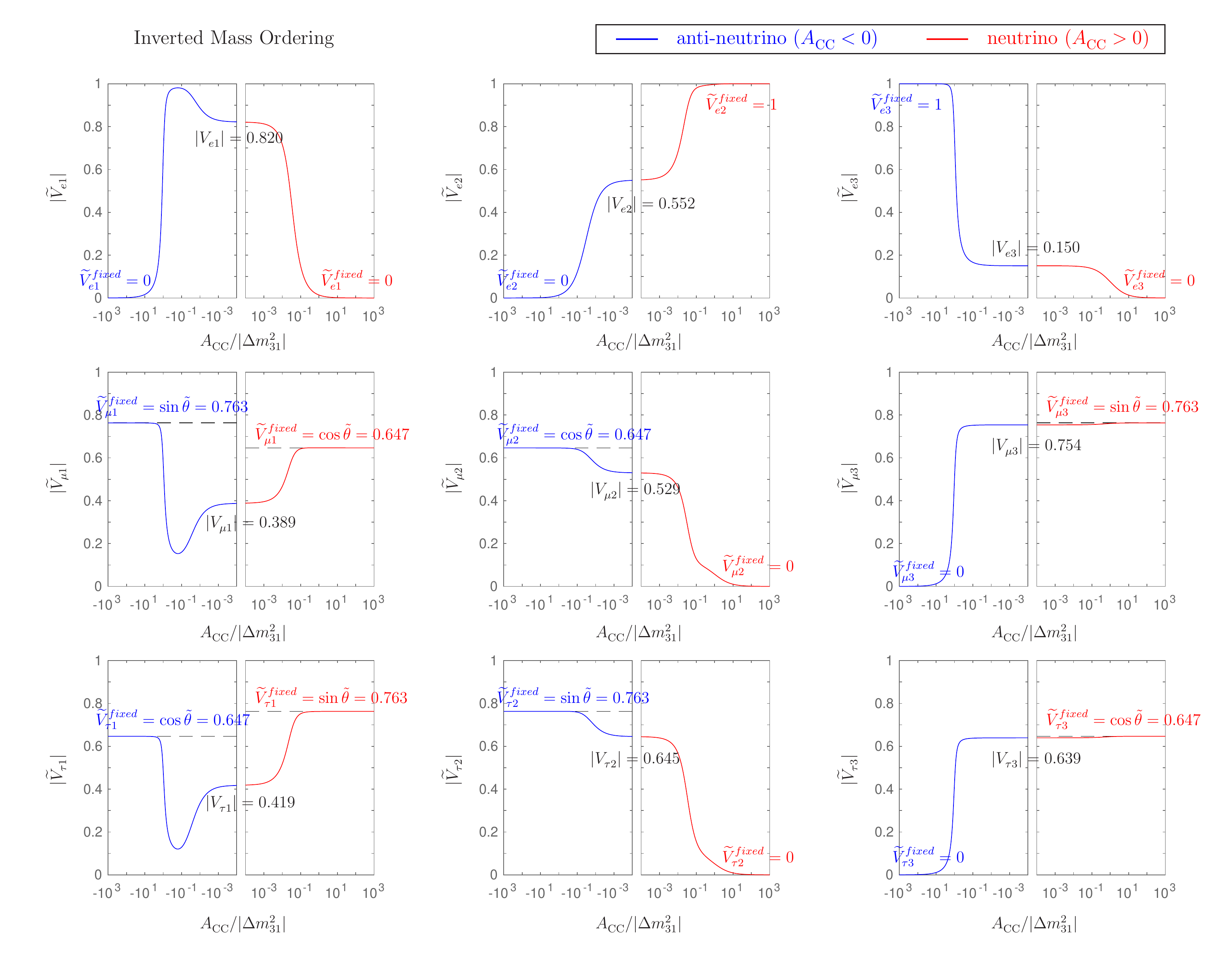}
\vspace{0cm}
\caption{The evolution of the absolute value of nine elements of the effective mixing matrix in matter $|\tilde{V}^{}_{\alpha i}|$ (for $\alpha = e, \mu, \tau$ and $i = 1, 2, 3$) with respect to the dimensionless ratio $A^{}_{\rm CC} / |\Delta m^{2}_{31}|$ in the inverted mass ordering case for both neutrinos (with $A^{}_{\rm CC} > 0$, red curves in each right half panel) and anti-neutrinos (with $A^{}_{\rm CC} < 0$, blue curves in each left half panel) , where the best-fit values of the mass-squared differences and the mixing parameters in Table. I have been input. Both the input values in vacuum and the fixed points in the limit $|A^{}_{\rm CC}| \gg |\Delta m^{2}_{31}|$ are given on the plots.}
\end{center}
\end{figure}

\begin{figure}
\begin{center}
\vspace{0cm}
\includegraphics[width=\textwidth]{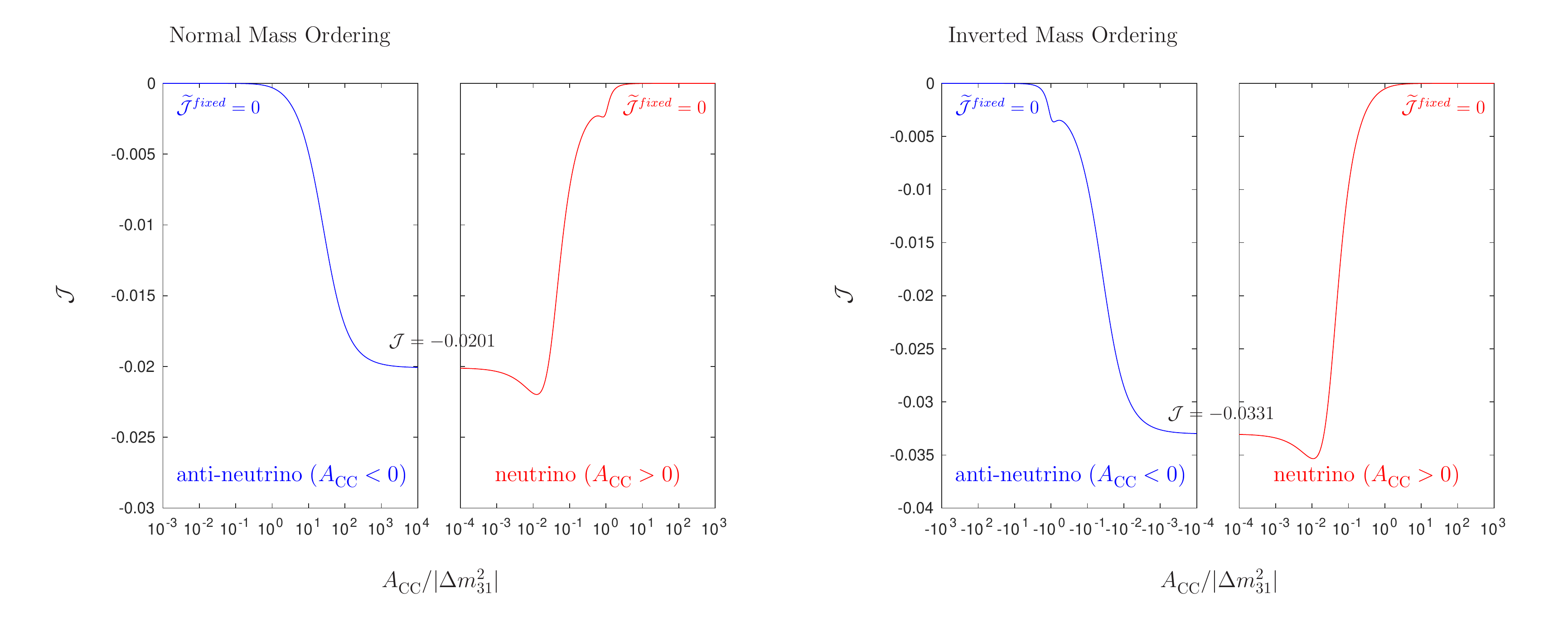}
\vspace{0cm}
\caption{The evolution of the effective Jarlskog invariant in matter $\tilde{\cal J}$ with respect to the dimensionless ratio $A^{}_{\rm CC} / |\Delta m^{2}_{31}|$ in the normal and the inverted mass ordering cases for both neutrinos (with $A^{}_{\rm CC} > 0$, red curves in the right half panel) and anti-neutrinos (with $A^{}_{\rm CC} < 0$, blue curves in the left half panel) , where the best-fit values of the mass-squared differences and the mixing parameters in Table. I have been input. Both the input values of $\cal J$ in vacuum and the fixed points $\tilde{\cal J}^{fixed}_{}$ in the limit $|A^{}_{\rm CC}| \gg |\Delta m^{2}_{31}|$ are given on the plots.}
\end{center}
\end{figure}

Note that, instead of ordering the eigenvalues according to their magnitude, we choose the order of $\tilde{\lambda}^{}_{i}$ in such a way that in the limit $|A^{}_{\rm CC}| \rightarrow 0$, the correct mass-squared differences in vacuum are obtained and the $i$th column of $\tilde{V}$ are corresponding eigenvectors of $\tilde{\lambda}^{}_{i}$. It's well known that in the standard three neutrinos framework there are two possible resonance regions (i.e., the solar resonance at around $A^{}_{\rm CC} \sim \Delta m^{2}_{21}$ and the atmospheric resonance at around $A^{}_{\rm CC} \sim \Delta m^{2}_{31}$) when studying the neutrino oscillation in matter. However, because the sign of $A^{}_{\rm CC}$ are different for neutrino or anti-neutrino oscillation and the sign of $\Delta m^{2}_{31}$ are different in the normal or inverted mass ordering case, above two resonance conditions are not always satisfied even if the magnitude of $A^{}_{\rm CC}$ could be carefully chosen. When passing through the resonance region, the related two eigenvalues ( as well as the corresponding two eigenvectors) ``exchange" their evolution behaviors. That explains the different patterns of the fixed points in different scenarios. Such a difference originates mainly from the fact that the resonances they experienced are different. To be specific, we list in Table II the different resonances neutrinos or anti-neutrinos with different mass orderings may experience together with the resulting pattern of the eigenvalues $\tilde{\lambda}^{}_{i}$ and the corresponding effective mixing matrix $\tilde{V}$ in the limit $|A^{}_{\rm CC}| \rightarrow \infty$ in different scenarios. Anyway, neither the ordering of the eigenvalues nor the omitted common terms would change the neutrino oscillation behaviors in matter which we will discuss in the next section.

One may clearly find from Figs. 1-5 that the evolutions of three effective neutrino masses $\tilde{m}^{}_{i}$, the absolute value of nine elements of the effective mixing matrix $|\tilde{V}^{}_{\alpha i}|$ as well as the effective Jarlskog parameter $\tilde{\cal J}$ actually follow a quite similar routine: in the vacuum-dominated region they slightly deviate from their vacuum inputs as $|A^{}_{\rm CC}|$ increases; when it enters the resonance region, these effective neutrino mass and mixing parameters receive dramatical corrections; and after the resonance region, the changes slowdown again and those effective parameters vary monotonically towards their fixed points in the matter-dominated region.

Although different eigenvalues $\tilde{\lambda}^{}_{i}$ and different elements of $\tilde{V}$  begin to approach their fixed points at different values of $A^{}_{\rm CC}$, we got an overall estimation from the numerical analysis that when $|A^{}_{\rm CC} / \Delta m^{2}_{31}| \gtrsim 10$ is satisfied, nine absolute differences $|\tilde{V}^{}_{\alpha i} - \tilde{V}^{fixed}_{\alpha i}|$ (for $\alpha = e, \mu, \tau$ and $i = 1, 2, 3$) and three relative differences $|\tilde{\lambda}^{}_{i} - \tilde{\lambda}^{fixed}_{i}| / \tilde{\lambda}^{fixed}_{i}$ (for $i = 1, 2, 3$) are all smaller than 0.01, and the absolute value of the effective Jarlskog $|\tilde{\cal J}| \lesssim 10^{-5}$ in all the four scenarios. We find that $|A^{}_{\rm CC} / \Delta m^{2}_{31}| \gtrsim 10$ could be regard as a good criterion of the ``matter-dominated region".

\section{Neutrino oscillation probabilities in the matter-dominated case}

If the matter density can be treated as a constant along the path neutrinos propagate, we can write down the neutrino oscillation probabilities in matter simply by replacing the neutrino mass-squared differences and the mixing matrix in the neutrino oscillation probabilities in vacuum with corresponding effective neutrino mass-squared differences and the effective mixing matrix in matter respectively. 
\begin{eqnarray}
P \; ( \stackrel{(-)}{\nu}^{}_{\alpha} \rightarrow  \stackrel{(-)}{\nu}^{}_{\beta} ) & = & \delta^{}_{\alpha \beta} - 4 \sum^{}_{j > i} {\rm Re} \left [ \tilde{V}^{}_{\alpha i} \tilde{V}^{}_{\beta j} \tilde{V}^{*}_{\alpha j} \tilde{V}^{*}_{\beta i} \right ] \sin^2 \tilde{\Delta}^{}_{ji} \pm 2 \sum^{}_{j > i} {\rm Im} \left [ \tilde{V}^{}_{\alpha i} \tilde{V}^{}_{\beta j} \tilde{V}^{*}_{\alpha j} \tilde{V}^{*}_{\beta i} \right ] \sin2\tilde{\Delta}^{}_{ji} \; , ~~
\end{eqnarray}
where $\tilde{\Delta}^{}_{ji} \equiv \Delta \tilde{m}^2_{ji} L / 4E$ with $\Delta \tilde{m}^2_{ji} \equiv \tilde{m}^{2}_{j} - \tilde{m}^{2}_{i} = 2 E ( \tilde{\lambda}^{}_{j} - \tilde{\lambda}^{}_{i} )$ being the effective neutrino mass-squared difference in matter. Here the Greek letters $\alpha$, $\beta$ are the flavor indices run over $e$, $\mu$, $\tau$, while the Latin letters $i$, $j$ are the indices of mass eigenstates run over $1$, $2$, $3$. And $E$ is the energy of the neutrino/anti-neutrino beam.

The neutrino oscillation probabilities to the second order of both $|\Delta m^{2}_{31} / A^{}_{\rm CC}|$ and $\Delta m^{2}_{21} / |A^{}_{\rm CC}|$ are derived in Appendix A. In the limit $|A^{}_{\rm CC}| \rightarrow \infty$, all terms proportional to $1 / A^{}_{\rm CC}$ become vanishing, then $\tilde{P}(\nu^{}_{\alpha} \rightarrow \nu^{}_{\beta})$ and $\tilde{P}(\bar{\nu}^{}_{\alpha} \rightarrow \bar{\nu}^{}_{\beta})$ approach to a same set of fixed values \footnote{This is in agreement with the vanishing of CP-violation in the limit $|A^{}_{\rm CC}| \rightarrow \infty$.}, and the neutrino/anti-neutrino oscillation probabilities in the matter-dominated region can be concisely expressed as
\begin{eqnarray}
\tilde{P}(\nu^{}_{e} \rightarrow \nu^{}_{e}) & \approx & \tilde{P}(\bar{\nu}^{}_{e} \rightarrow \bar{\nu}^{}_{e}) \; \approx \; 1 \; , \nonumber\\[1mm]
\tilde{P}(\nu^{}_{e} \rightarrow \nu^{}_{\mu}) & \approx & \tilde{P}(\bar{\nu}^{}_{e} \rightarrow \bar{\nu}^{}_{\mu}) \; \approx \; 0 \; , \nonumber\\[1mm]
\tilde{P}(\nu^{}_{e} \rightarrow \nu^{}_{\tau}) & \approx & \tilde{P}(\bar{\nu}^{}_{e} \rightarrow \bar{\nu}^{}_{\tau}) \; \approx \; 0 \; , \nonumber\\[1mm]
\tilde{P}(\nu^{}_{\mu} \rightarrow \nu^{}_{e}) & \approx & \tilde{P}(\bar{\nu}^{}_{\mu} \rightarrow \bar{\nu}^{}_{e}) \; \approx \; 0 \; , \nonumber\\[1mm]
\tilde{P}(\nu^{}_{\mu} \rightarrow \nu^{}_{\mu}) & \approx & \tilde{P}(\bar{\nu}^{}_{\mu} \rightarrow \bar{\nu}^{}_{\mu}) \; \approx \; 1 - \sin^2 2\tilde{\theta} \sin^2 \frac{\Delta \tilde{m}^{2}_{32} L}{4 E} \; , \nonumber\\[1mm]
\tilde{P}(\nu^{}_{\mu} \rightarrow \nu^{}_{\tau}) & \approx & \tilde{P}(\bar{\nu}^{}_{\mu} \rightarrow \bar{\nu}^{}_{\tau}) \; \approx \; \sin^2 2\tilde{\theta} \sin^2 \frac{\Delta \tilde{m}^{2}_{32} L}{4 E} \; , \nonumber\\[1mm]
\tilde{P}(\nu^{}_{\tau} \rightarrow \nu^{}_{e}) & \approx & \tilde{P}(\bar{\nu}^{}_{\tau} \rightarrow \bar{\nu}^{}_{e}) \; \approx \; 0 \; , \nonumber\\[1mm]
\tilde{P}(\nu^{}_{\tau} \rightarrow \nu^{}_{\mu}) & \approx & \tilde{P}(\bar{\nu}^{}_{\tau} \rightarrow \bar{\nu}^{}_{\mu}) \; \approx \; \sin^2 2\tilde{\theta} \sin^2 \frac{\Delta \tilde{m}^{2}_{32} L}{4 E} \; , \nonumber\\[1mm]
\tilde{P}(\nu^{}_{\tau} \rightarrow \nu^{}_{\tau}) & \approx & \tilde{P}(\bar{\nu}^{}_{\tau} \rightarrow \bar{\nu}^{}_{\tau}) \; \approx \; 1 - \sin^2 2\tilde{\theta} \sin^2 \frac{\Delta \tilde{m}^{2}_{32} L}{4 E} \; ,
\end{eqnarray}
where
\begin{eqnarray}
\Delta \tilde{m}^{2}_{32} & \approx & \left [ \Delta m^{2}_{21} \left ( | V^{}_{\tau 2} |^{2}_{} - | V^{}_{\mu 2} |^{2}_{} \right ) + \Delta m^{2}_{31} \left ( | V^{}_{\tau 3} |^{2}_{} - | V^{}_{\mu 3} |^{2}_{} \right ) \right ] \cos2\tilde{\theta} \nonumber\\
& & + 2 | \Delta m^{2}_{21} V^{}_{\mu 2} V^{*}_{\tau 2}  + \Delta m^{2}_{31} V^{}_{\mu 3} V^{*}_{\tau 3} | \sin2\tilde{\theta} \nonumber\\[2mm]
& = & \pm \sqrt{\left ( \Delta m^{2}_{31} c^{2}_{13} - \Delta m^{2}_{21} ( c^{2}_{12} -s^{2}_{12} s^{2}_{13} ) \right )^2 + \left( \Delta m^{2}_{21} \sin2\theta^{}_{12} s^{}_{13} \right )^2} \; .
\end{eqnarray}
Here $\Delta \tilde{m}^{2}_{32}$ has the same sign as $\Delta m^{2}_{31}$. Again, taking into account the strong hierarchy of $\Delta m^{2}_{21} \ll |\Delta m^{2}_{31}|$ and the smallness of $s^{}_{13}$, we can then obtain that the effective mass-squared difference $\Delta \tilde{m}^{2}_{32} \approx \Delta m^{2}_{32}$ (or $\Delta m^{2}_{31}$) \footnote{In our numerical analysis, we have $\Delta \tilde{m}^{2}_{32} = 2.349 \times10^{-3} {\rm eV}^2$ together with $\Delta m^{2}_{32} = 2.3772 \times10^{-3} {\rm eV}^2$ in the normal mass ordering case, and $\Delta \tilde{m}^{2}_{32} = -2.501 \times10^{-3} {\rm eV}^2$ together with $\Delta m^{2}_{32} = -2.5859 \times10^{-3} {\rm eV}^2$ in the inverted mass ordering case.} together with $\tilde{\theta} \approx \theta^{}_{23}$. 

These analytical approximations give us a clear picture of neutrino oscillation in the matter-dominated region: $\nu^{}_{e}$ are decoupled (due to its intense charged-current interaction with electrons in the medium), while oscillation can still happened between $\nu^{}_{\mu}$ and $\nu^{}_{\tau}$ \footnote{This is in agreement with the near degeneracy of $\tilde{\lambda}^{}_{2}$ and $\tilde{\lambda}^{}_{3}$ in the limit $|A^{}_{\rm CC}| \rightarrow \infty$.}. This two-flavor oscillation can be described by one effective mixing angle $\tilde{\theta}$ and the effective mass-squared difference $\Delta \tilde{m}^{2}_{32}$ whose expressions are given in Eqs. (7) and (10) respectively. Note that, both the oscillation parameters are independent of $A^{}_{\rm CC}$ and can be easily calculated once the neutrino oscillation parameters in vacuum are well determined. It means as long as the ``matter-dominated'' condition is satisfied, above simple formulas are applicable no matter how the matter density varies along the path, and the resulting conversion probability between $\nu^{}_{\mu}$ and $\nu^{}_{\tau}$ is just a simple function of $L / E$.  

\begin{figure}
\begin{center}
\vspace{0cm}
\includegraphics[width=\textwidth]{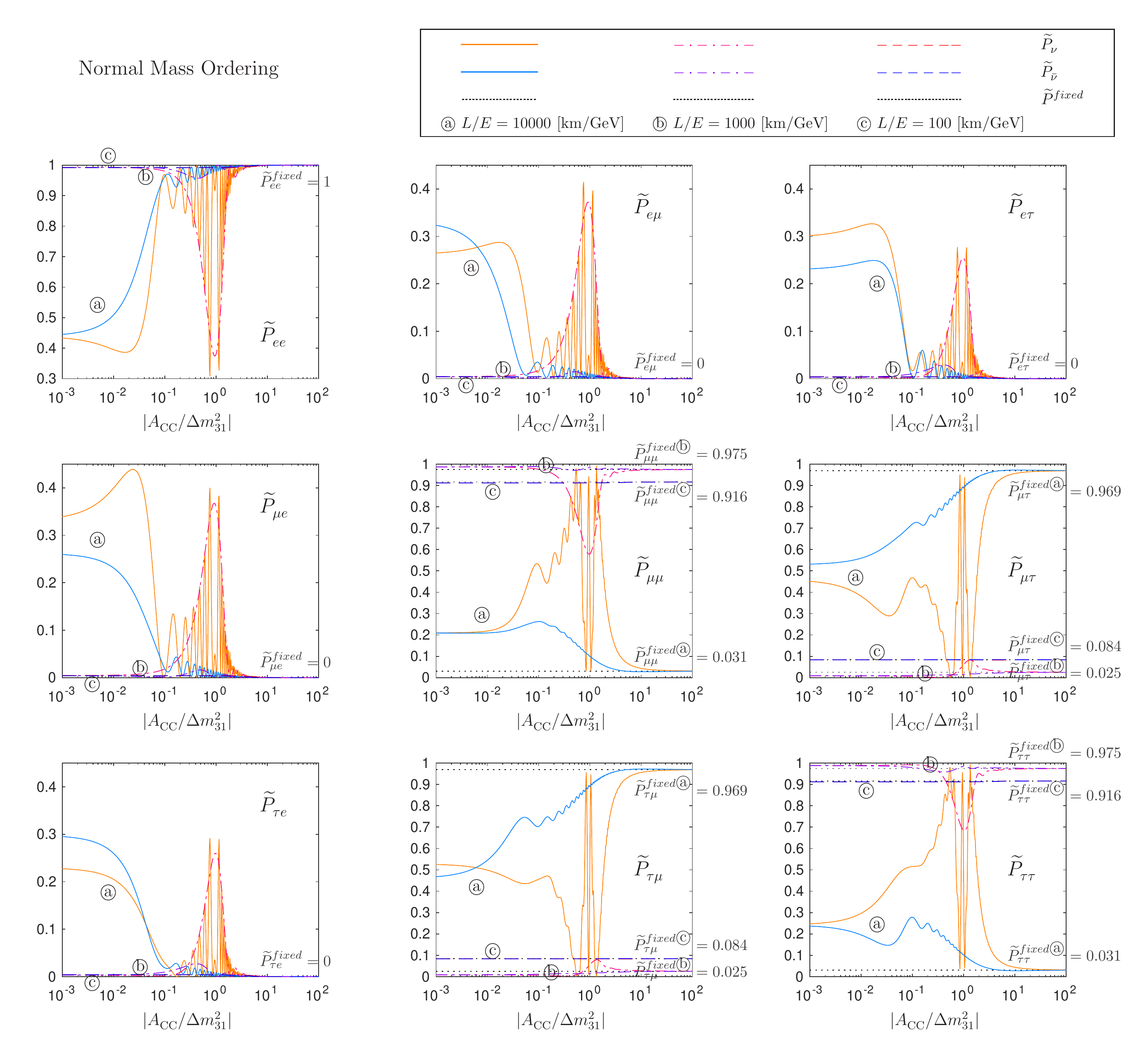}
\vspace{0cm}
\caption{For three different values of $L / E$, the evolution of the oscillation probabilities in matter $\tilde{P}^{}_{\alpha\beta}$ with respect to the dimensionless ratio $|A^{}_{\rm CC} / \Delta m^{2}_{31}|$ in the normal mass ordering case for both neutrinos and anti-neutrinos are shown in this figure, where the best-fit values of the mass-squared differences and the mixing parameters in Table. I have been input. The fixed points of these probabilities in the limit $|A^{}_{\rm CC}| \gg |\Delta m^{2}_{31}|$ for different $L/E$ are given on the plots.}
\end{center}
\end{figure}

\begin{figure}
\begin{center}
\vspace{0cm}
\includegraphics[width=\textwidth]{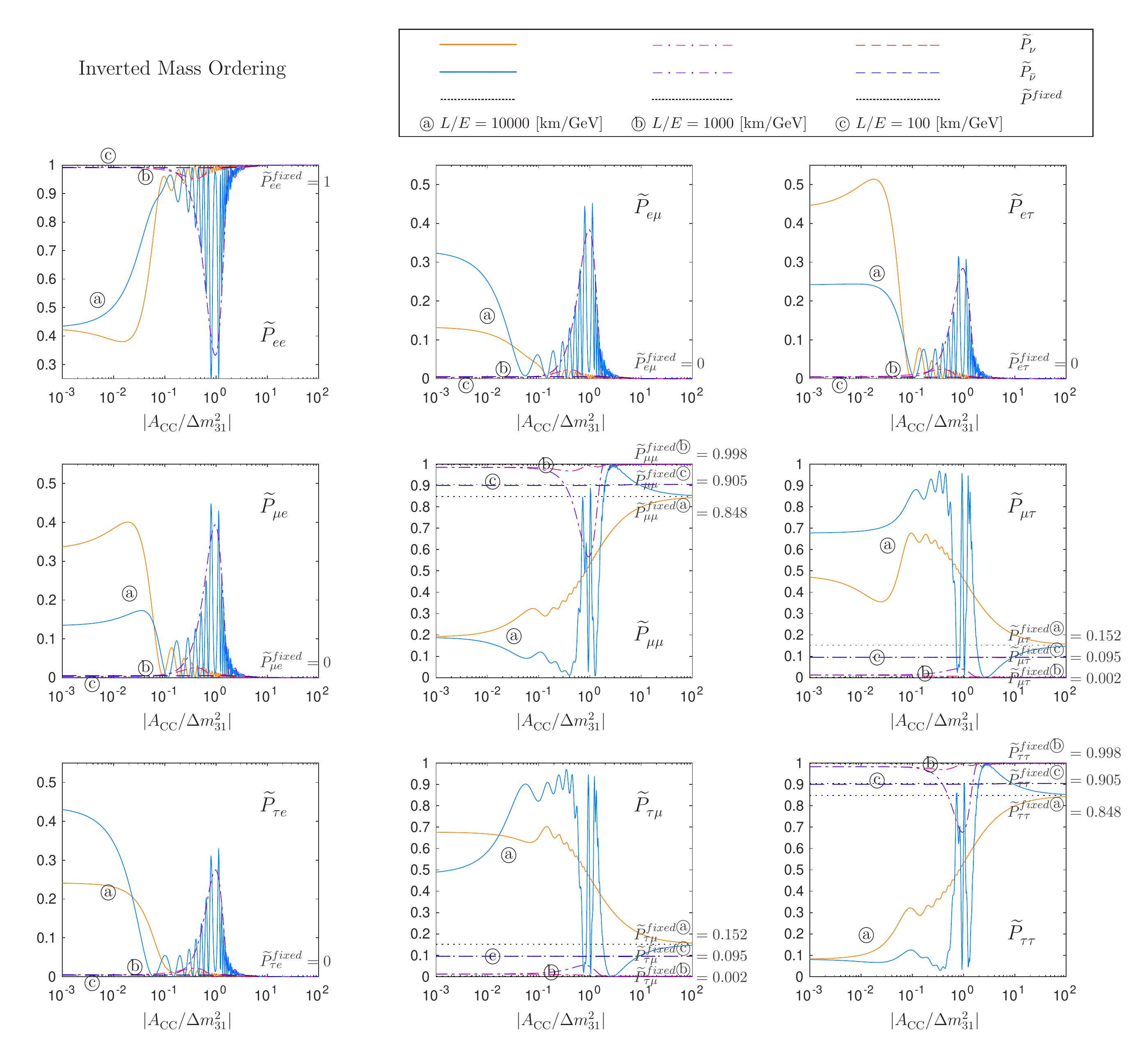}
\vspace{0cm}
\caption{For three different values of $L / E$, the evolution of the oscillation probabilities in matter $\tilde{P}^{}_{\alpha\beta}$ with respect to the dimensionless ratio $|A^{}_{\rm CC} / \Delta m^{2}_{31}|$ in the inverted mass ordering case for both neutrinos and anti-neutrinos are shown in this figure, where the best-fit values of the mass-squared differences and the mixing parameters in Table. I have been input. The fixed points of these probabilities in the limit $|A^{}_{\rm CC}| \gg |\Delta m^{2}_{31}|$ for different $L/E$ are given on the plots.}
\end{center}
\end{figure}

To help understand the general picture of neutrino/anti-neutrino oscillation in matter, we choose three different values of $L / E = 10000, 1000, 100$ [km/GeV] and illustrate in Figs. 6 and 7 the evolution of the neutrino/anti-neutrino oscillation probabilities in matter $\tilde{P}^{}_{\alpha\beta}$ with respect to the dimensionless ratio $A^{}_{\rm CC} / |\Delta m^{2}_{31}|$ in both the normal and the inverted mass ordering cases.

Analogous to those effective neutrino mass and mixing parameters we discussed in Sec. II, one can clearly distinguish the ``vacuum-dominated", ``resonance",  and ``matter-dominated" three different regions in Figs. 6 and 7. In the vacuum-dominated region ($|A^{}_{\rm CC}| \ll \Delta m^{2}_{21}$) the oscillation probabilities receive relatively mild corrections with respect to their vacuum inputs.  At around $|A^{}_{\rm CC}| \sim \Delta m^{2}_{21}$ (i.e., the solar resonance region) corrections to the neutrino oscillation probabilities become significant, while at around $|A^{}_{\rm CC}| \sim \Delta m^{2}_{31}$ (i.e., the atmospheric resonance region)  the neutrino oscillation probabilities in the normal ordering case and the anti-neutrino oscillation probabilities in the inverted mass ordering case receive dramatical corrections. Then after the resonance region, both the neutrino and the anti-neutrino oscillation probabilities quickly evolve to the same set of fixed points ($\tilde{P}^{}_{\alpha\beta} \simeq \tilde{P}^{}_{\bar{\alpha} \bar{\beta}} \simeq \tilde{P}^{fixed}_{\alpha\beta}$) in the matter-dominated region which have been well predicted in Eq. (9), and the CP violation tend to vanish in this region. Among the nine $\tilde{P}^{fixed}_{\alpha\beta}$, we have $\tilde{P}^{fixed}_{ee} = 1$ and $\tilde{P}^{fixed}_{e\mu} = \tilde{P}^{fixed}_{e\tau} = \tilde{P}^{fixed}_{\mu e} = \tilde{P}^{fixed}_{\tau e} = 0$, which tells us that $\nu^{}_{e} / \bar{\nu}^{}_{e}$ decouples from the other flavors in the  matter-dominated case. Although the oscillation probabilities between $\nu^{}_{\mu}$ and $\nu^{}_{\tau}$ are approximately independent of $A^{}_{\rm CC}$ in the matter-dominated region, $\tilde{P}^{}_{\mu\mu}$ and $\tilde{P}^{}_{\tau\tau}$ ($\tilde{P}^{}_{\mu\tau}$ and $\tilde{P}^{}_{\tau\mu}$) change periodically between 1 and $1 - \sin2\tilde{\theta}$ (0 and $\sin2\tilde{\theta}$) as the variation of $L / E$. Note that,  $\sin2\tilde{\theta}$ is actually close to 1, which means if $L$ and $E$ are properly chosen, a simple but significant two-flavor oscillation between $\nu^{}_{\mu}$ and $\nu^{}_{\tau}$ can be observed in the matter-dominated case.

\begin{figure}
\begin{center}
\vspace{0cm}
\includegraphics[width=\textwidth]{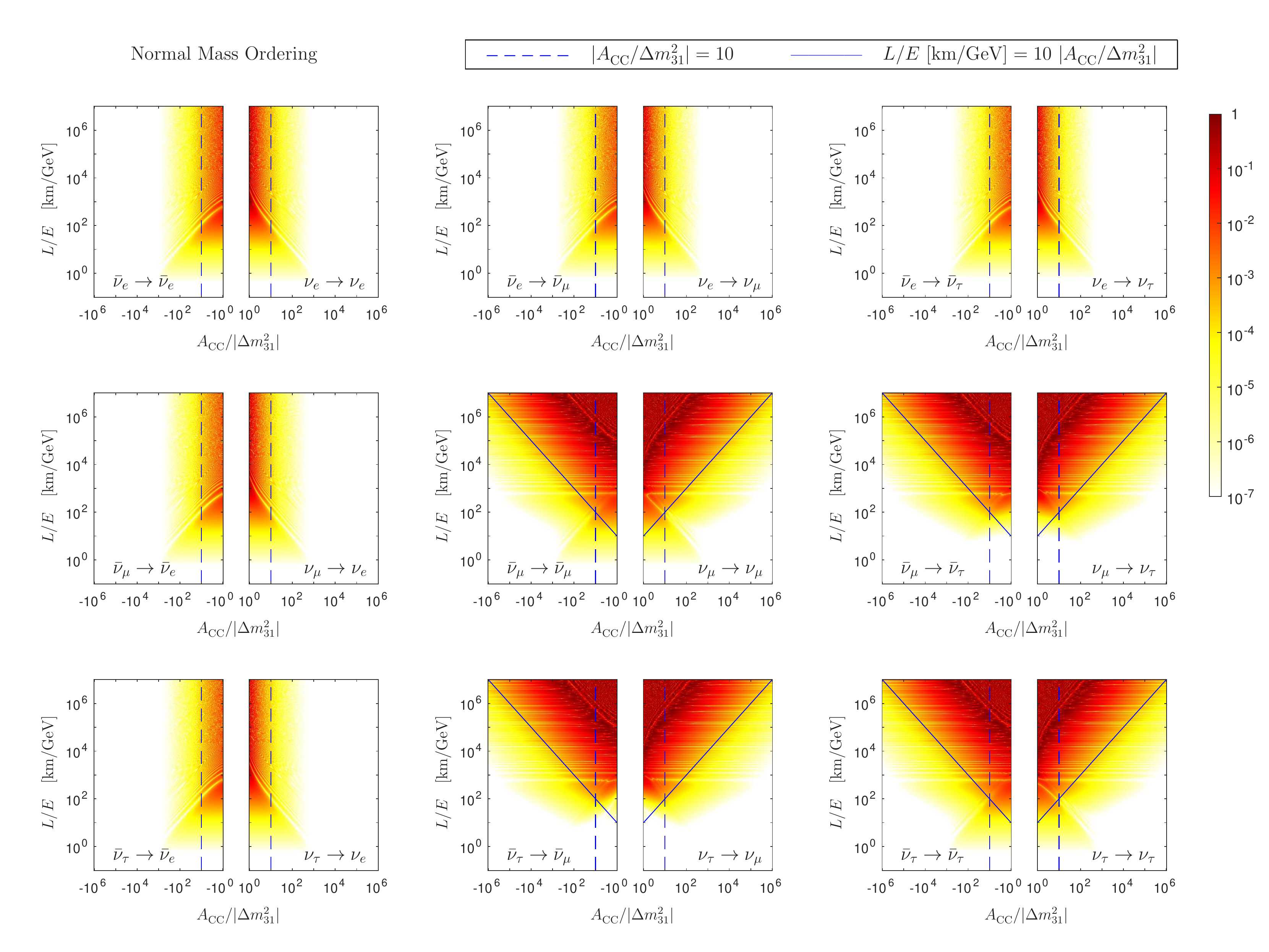}
\vspace{0cm}
\caption{These color images display the absolute errors $|\tilde{P}^{}_{\alpha\beta} - \tilde{P}^{fixed}_{\alpha\beta}|$ for neutrinos in each right half panel (with $A^{}_{\rm CC} > 0$) and $|\tilde{P}^{}_{\bar{\alpha}\bar{\beta}} - \tilde{P}^{fixed}_{\alpha\beta}|$ for anti-neutrinos in each left half panel (with $A^{}_{\rm CC} < 0$) in the normal mass ordering case, where the best-fit values of the mass-squared differences and the mixing parameters in Table. I have been input. In these images, $\tilde{P}^{}_{\alpha\beta}$ or $\tilde{P}^{}_{\bar{\alpha}\bar{\beta}}$ are calculated by numerically diagonalizing the effective Hamiltonian $\tilde{\cal H}$ in matter.}
\end{center}
\end{figure}

\begin{figure}
\begin{center}
\vspace{0cm} 
\includegraphics[width=\textwidth]{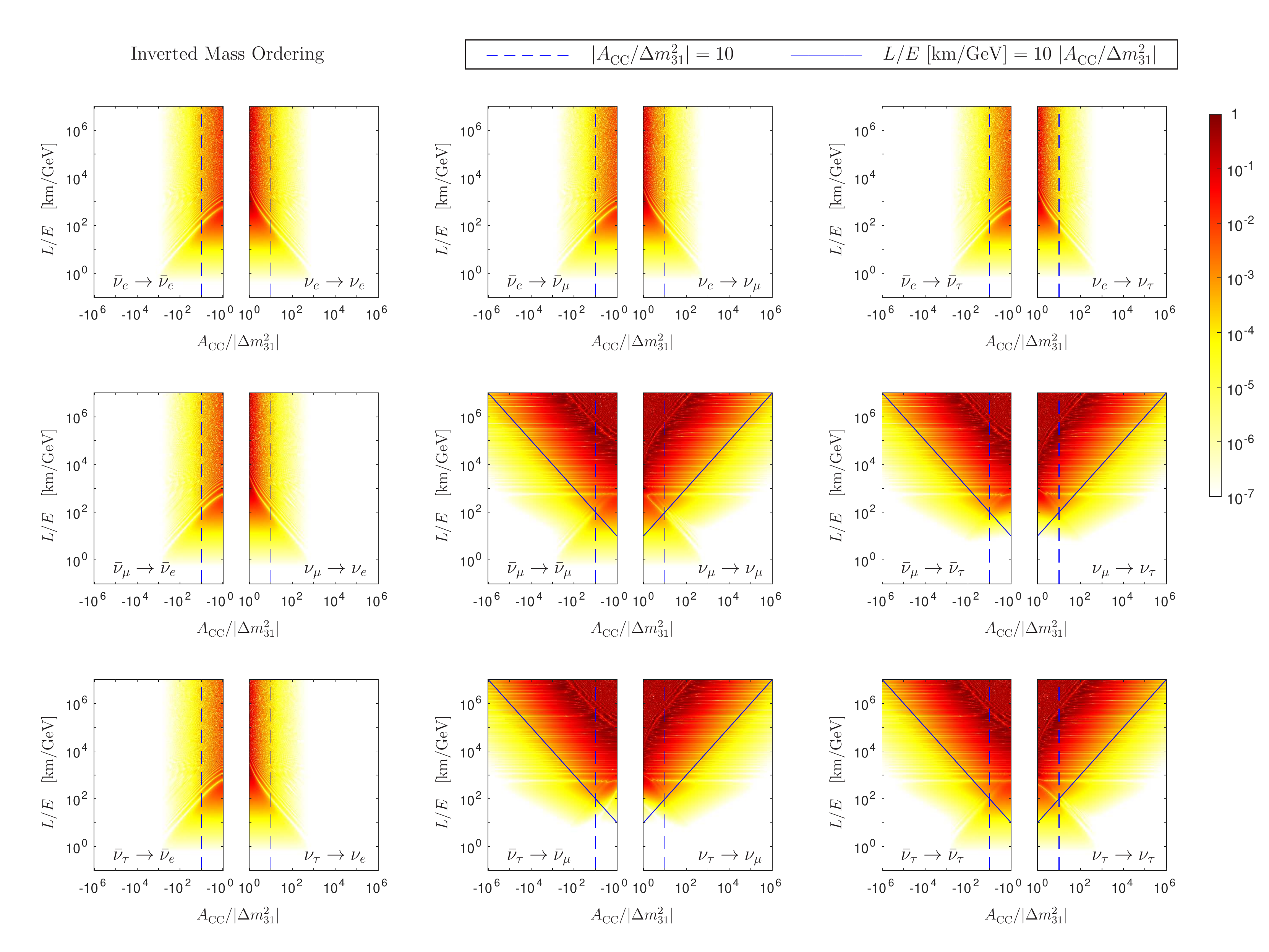}
\vspace{0cm}
\caption{These color images display the absolute errors $|\tilde{P}^{}_{\alpha\beta} - \tilde{P}^{fixed}_{\alpha\beta}|$ for neutrinos in each right half panel (with $A^{}_{\rm CC} > 0$) and $|\tilde{P}^{}_{\bar{\alpha}\bar{\beta}} - \tilde{P}^{fixed}_{\alpha\beta}|$ for anti-neutrinos in each left half panel (with $A^{}_{\rm CC} < 0$) in the inverted mass ordering case, where the best-fit values of the mass-squared differences and the mixing parameters in Table. I have been input. In these images, $\tilde{P}^{}_{\alpha\beta}$ or $\tilde{P}^{}_{\bar{\alpha}\bar{\beta}}$ are calculated by numerically diagonalizing the effective Hamiltonian $\tilde{\cal H}$ in matter.}
\end{center}
\end{figure}

Before ending this section, we would like to test the accuracy of the formulas given in Eq. (9) and discuss the valid region of these formulas. Figures 8 and 9 show the absolute errors $|\tilde{P}^{}_{\alpha\beta} - \tilde{P}^{fixed}_{\alpha\beta}|$ of neutrino/anti-neutrino oscillation probabilities in both the normal and the inverted mass ordering cases, where $\tilde{P}^{fixed}_{\alpha\beta}$ is calculated using Eq. (9) and $\tilde{P}^{}_{\alpha\beta}$ is numerically calculated without any approximation.
In previous discussion, we have employed $|A^{}_{\rm CC} / \Delta m^{2}_{31}| \gtrsim 10$ as the criterion of the matter-dominated condition, i.e., the matter term ${\cal H}'$ is at least an order larger than the vacuum Hamiltonian ${\cal H}$. As we can see from Figs. 8 and 9, under this criterion, the differences of $\tilde{P}^{}_{ee}$, $\tilde{P}^{}_{e\mu}$, $\tilde{P}^{}_{e\tau}$, $\tilde{P}^{}_{\mu e}$ and $\tilde{P}^{}_{\tau e}$ with respect to their fixed points (1 or 0) are all smaller than $10^{-4}$. If a more strict criterion $|A^{}_{\rm CC} / \Delta m^{2}_{31}| \gtrsim 100$ is adopted, the absolute error of these oscillation probabilities related to electron flavor would be smaller than $10^{-7}$. And as one can infer from Eq. (A15), the absolute errors would fall quadratically with the increase of $|A^{}_{\rm CC}|$. We can then safely make the conclusion that both $\nu^{}_{e}$ and $\bar{\nu}^{}_{e}$ are decoupled in the matter-dominated case.
On the other hand, in addition to the dependence on the matter parameter $A^{}_{\rm CC}$, the accuracy of the oscillation probabilities $\tilde{P}^{}_{\mu\mu}$, $\tilde{P}^{}_{\tau\tau}$, $\tilde{P}^{}_{\mu\tau}$ and $\tilde{P}^{}_{\tau\mu}$ which describe the  remaining oscillation between $\nu^{}_{\mu}$ and $\nu^{}_{\tau}$ in dense matter depend also crucially on the ratio $L / E$. If both the conditions $|A^{}_{\rm CC} / \Delta m^{2}_{31}| \gtrsim 10$ and $L / E ~ [{\rm km} / {\rm GeV}] \lesssim 10 |A^{}_{\rm CC} / \Delta m^{2}_{31}|$ are satisfied, the absolute error of these four probabilities are all smaller than $10^{-3}$. And if the more strict constraint $L / E ~ [{\rm km} / {\rm GeV}] \lesssim |A^{}_{\rm CC} / \Delta m^{2}_{31}|$ together with $|A^{}_{\rm CC} / \Delta m^{2}_{31}| \gtrsim 10$ are imposed on, the accuracy of the order $10^{-5}$ or better can be obtained. The reason for this additional criterion is that the first order correction to the effective mass-squared difference $\Delta \tilde{m}^{2}_{32}$ is proportional to $|\Delta m^{2}_{31} / A^{}_{\rm CC}|$ as on can see in Eq. (A16). In the case $L / E ~ [{\rm km} / {\rm GeV}] \gtrsim 10 |A^{}_{\rm CC} / \Delta m^{2}_{31}|$ this correction to the oscillation frequency is significant enough and should not be ignored. In this case one may calculate $\Delta \tilde{m}^{2}_{32}$ using Eq. (A16) instead of Eq. (10) to further improve the accuracy of Eq. (9). Also note that, when a realistic experiment is discussed especially for those with large $\Delta \tilde{m}^{2}_{32} L /4 E$, the energy resolution must be taken into consideration.

\section{Outlook}

As the ending section of this manuscript, it is interesting to ask under what circumstances these studies of neutrino oscillation in dense matter will be applied.
Here we bring our embryo thoughts by making a very bold comparison of the oscillation behaviors between neutrinos passing through the Earth and passing through a typical white dwarf.

\begin{figure}[!h]
\begin{center}
\vspace{0cm}
\includegraphics[width=\textwidth]{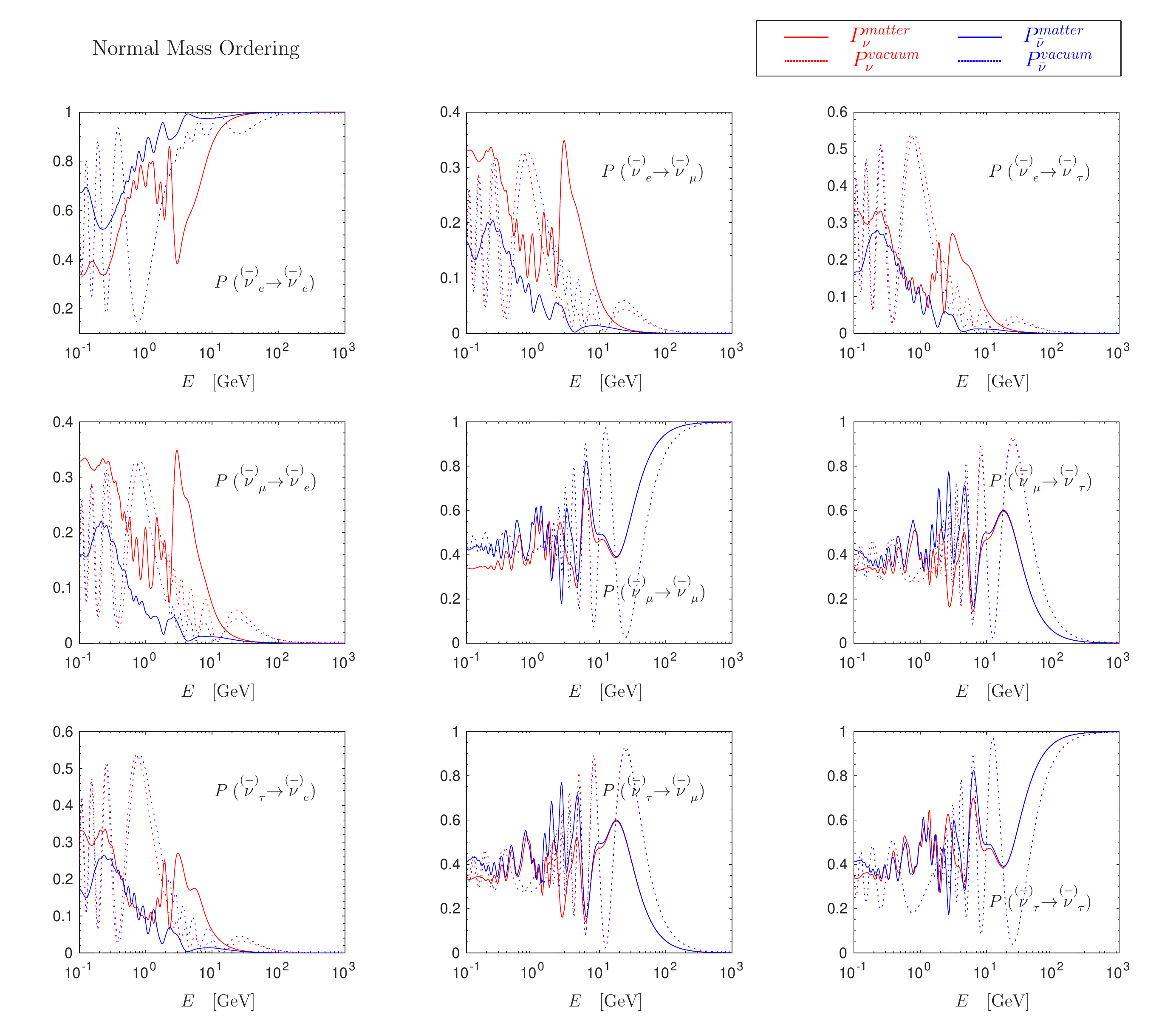}
\vspace{0cm}
\caption{The comparison of the neutrino (anti-neutrino) oscillation probabilities with or without the matter effect as a neutrino (anti-neutrino) beam of energy $E$ go through the Earth along the diameter, where the normal neutrino mass ordering is assumed and the best-fit values of the mass-squared differences and the mixing parameters in Table. I have been input. The fixed points of the probabilities in the limit $|A^{}_{\rm CC}| \gg |\Delta m^{2}_{31}|$ given by Eq. (9) (dashed lines) are also plotted in this figure for comparison. Note that all the probabilities are averaged over a Gaussian energy resolution of $5\%$.}
\end{center}
\end{figure}

\begin{figure}[!h]
\begin{center}
\vspace{0cm}
\includegraphics[width=\textwidth]{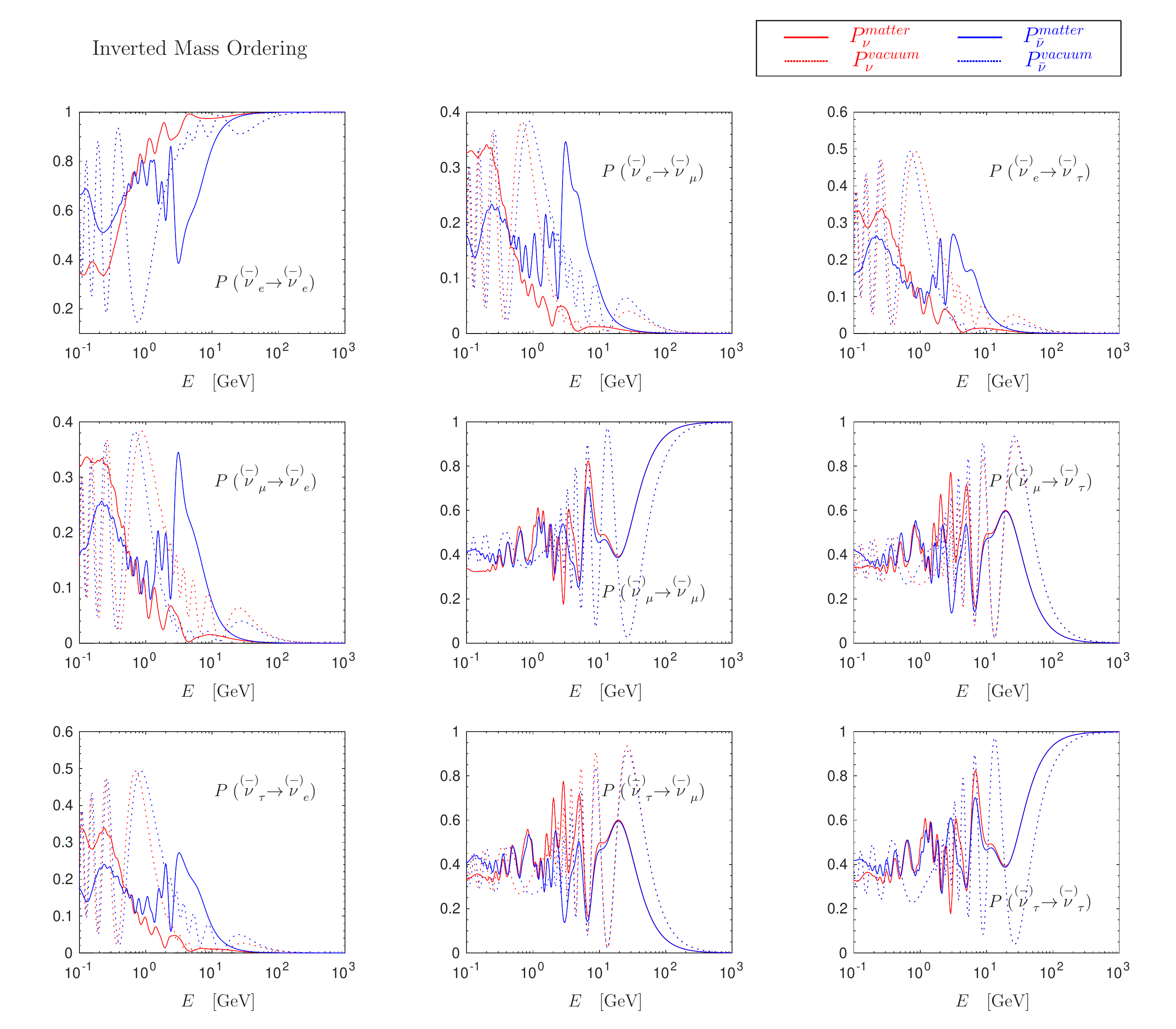}
\vspace{0cm}
\caption{The comparison of the neutrino (anti-neutrino) oscillation probabilities with or without the matter effect as a neutrino (anti-neutrino) beam of energy $E$ go through the Earth along the diameter, where the inverted neutrino mass ordering is assumed and the best-fit values of the mass-squared differences and the mixing parameters in Table. I have been input. The fixed points of the probabilities in the limit $|A^{}_{\rm CC}| \gg |\Delta m^{2}_{31}|$ given by Eq. (9) (dashed lines) are also plotted in this figure for comparison. Note that all the probabilities are averaged over a Gaussian energy resolution of $5\%$.}
\end{center}
\end{figure}

Figures 10 and 11 show the variations of oscillation probabilities as functions of neutrino/anti-neutrino energy $E$, when the neutrino/anti-neutrino beam go through the Earth along the diameter.  Note that, instead of the more accurate PREM model of the Earth \cite{PREM}, we adopted here a simpler two layer mantle-core model \cite{Stacey, Maris:1997nk} \footnote{Our numerical analysis show that there is no discernible difference between results using these two different Earth reference models in both the vacuum-dominated and the matter-dominated regions.}. The Earth radius in this mantle-core model is $R = 6371 ~ {\rm km}$, of which the core has a radius of $R^{}_{c} = 3458.7 ~ {\rm km}$ with the average matter density $\bar{\rho}^{}_{c} \simeq 11.5 ~ {\rm g} / {\rm cm}^{3}$ and the average electron fraction $Y^{c}_{e} \simeq 0.467$, while the Earth mantle has an approximately depth of $R^{}_{m} = 2885.3 ~ {\rm km}$ with the average matter density $\bar{\rho}^{}_{m} \simeq 4.5 ~ {\rm g} / {\rm cm}^{3}$ and the average electron fraction $Y^{m}_{e} \simeq 0.49$. 

We can find from these figures, when $E \gtrsim 100 ~{\rm GeV}$, the criterion of ``matter-dominated'' is satisfied and the probabilities of neutrino oscillation and anti-neutrino oscillation in this region are approximately equal and can be well predicted by Eq. (9). However in the case of our Earth, $L / E$ is pretty small in the matter-dominated region, then all the disappearance probabilities are approximately equal to 1 while all the appearance probabilities are approximately equal to 0. The neutrinos/anti-neutrinos of all three flavors are decoupled in this case. Also it's worth mentioning that, since the neutrino-nucleon cross section of neutrinos increase with increasing energy \cite{Gandhi:1995tf, Gandhi:1998ri}, at such high energies the Earth becomes opaque to neutrinos and the neutrino flux gets attenuated (for more details, see discussions in e.g., \cite{Naumov:1998sf, Palomares-Ruiz:2015mka, Vincent:2017svp, Donini:2018tsg}). In the case of neutral-current interaction neutrinos are degraded in energy, and in the case of charged-current interaction neutrinos are absorbed. The attenuation becomes more important than the usual matter effect in such high energy region and should not be ignored.

However, if we can do the same measurements on a white dwarf whose volume is comparable to that of the Earth but mass is comparable to that of the Sun, things could have been very different. Figures 12 and 13 show the variations of oscillation probabilities as functions of neutrino/anti-neutrino energy $E$, when the neutrino/anti-neutrino beam passing through a typical white dwarf \cite{Shapiro:1983du, MaxCamenzind} along its diameter. The corresponding oscillation probabilities in vacuum are also presented in these plots using dotted lines for comparison. Again, all the probabilities are averaged over a Gaussian energy resolution of $5\%$. The white dwarf is an excellent choice for this thought experiment. On one hand a white dwarf is very dense can give rise to significant matter effect, and on the other hand the material in a white dwarf no longer undergoes fusion reactions which means it does not radiate large amount of neutrinos on its own. In our analysis, the mass $M \sim 0.7 M^{}_{\odot}$ (with $M^{}_{\odot}$ being the mass of the Sun), the radius $R \sim 10^{4} ~{\rm km}$, an uniform density $\rho \sim 2 \times 10^{6} ~ {\rm g} / {\rm cm}^{3}$ or equivalently an uniform electron number density $n^{}_{e} \sim 6 \times 10^{29} ~ {\rm cm}^{-3} \sim 10^{6} N^{}_{\rm A} ~{\rm cm}^{-3}$ (with $N^{}_{\rm A}$ being the Avogadro's number) are assumed as the properties of this white dwarf. 

Due to the extremely high density, neutrino oscillation experiences the resonances and then enter the matter-dominated region at very low energies (below MeV). One may clearly see from Figs. 12 and 13 that at around $E \sim 0.4 ~{\rm keV}$ (the solar resonance where $|A^{}_{\rm CC}| \simeq \Delta m^{2}_{21}$) the oscillation probabilities start to markedly differ from the vacuum oscillation probabilities and change towards their fixed points. For neutrino oscillation in the normal mass ordering case or anti-neutrino oscillation in the inverted mass ordering case, there is a significant resonance hump at around $E \sim 20 ~{\rm keV}$ (the atmospheric resonance where $|A^{}_{\rm CC}| \simeq |\Delta m^{2}_{31}|$). After that, at around $E \sim 0.2 ~{\rm MeV}$ (where $|A^{}_{\rm CC}|/|\Delta m^{2}_{31}| \simeq 10$), it enters the matter-dominated region. In our analysis, the neutrino/anti-neutrino oscillation probabilities in this region are all in perfect agreement with the predictions of Eq. (9) if the same energy resolution is taken into account. In the energy range shown in these two figures, $L / E$ is extremely large, the oscillatory frequencies are all  extremely high, therefore only the average oscillatory magnitude can be observed, which is a constant and is markedly different from the vacuum probabilities in the matter-dominated case. 

It's worth mentioning that, in the low energy region, the oscillatory frequency $\Delta \tilde{m}^{2}_{ji} L / 4 E$ could be high. In this case neutrinos undergo very quick oscillations which can not actually be observed due to the finite energy resolution of the detectors. In our numerical analysis presented in Figs. 10-13, all the probabilities are averaged over a Gaussian energy resolution of $5\%$ (which can be achieved by the upcoming neutrino experiments, such as JUNO \cite{An:2015jdp}, at the MeV energy range) to mimic the working of the detector on one hand and uncover features hidden in these fast oscillations on the other. Our numerical analysis also show that even if we choose a worse energy resolution of $15\%$, the intriguing features discussed above can still be well recognized, since we are looking for the resonance hump and the deviation of the average oscillatory magnitude after neutrinos passing through the white dwarf instead of looking for the oscillation behavior itself. However if we want to trace the remaining oscillation between $\nu^{}_{\mu}$ and $\nu^{}_{\tau}$ in this dense matter at a much higher energy range, a good energy resolution could be crucially important. 

\begin{figure}[!h]
\begin{center}
\vspace{0cm}
\includegraphics[width=\textwidth]{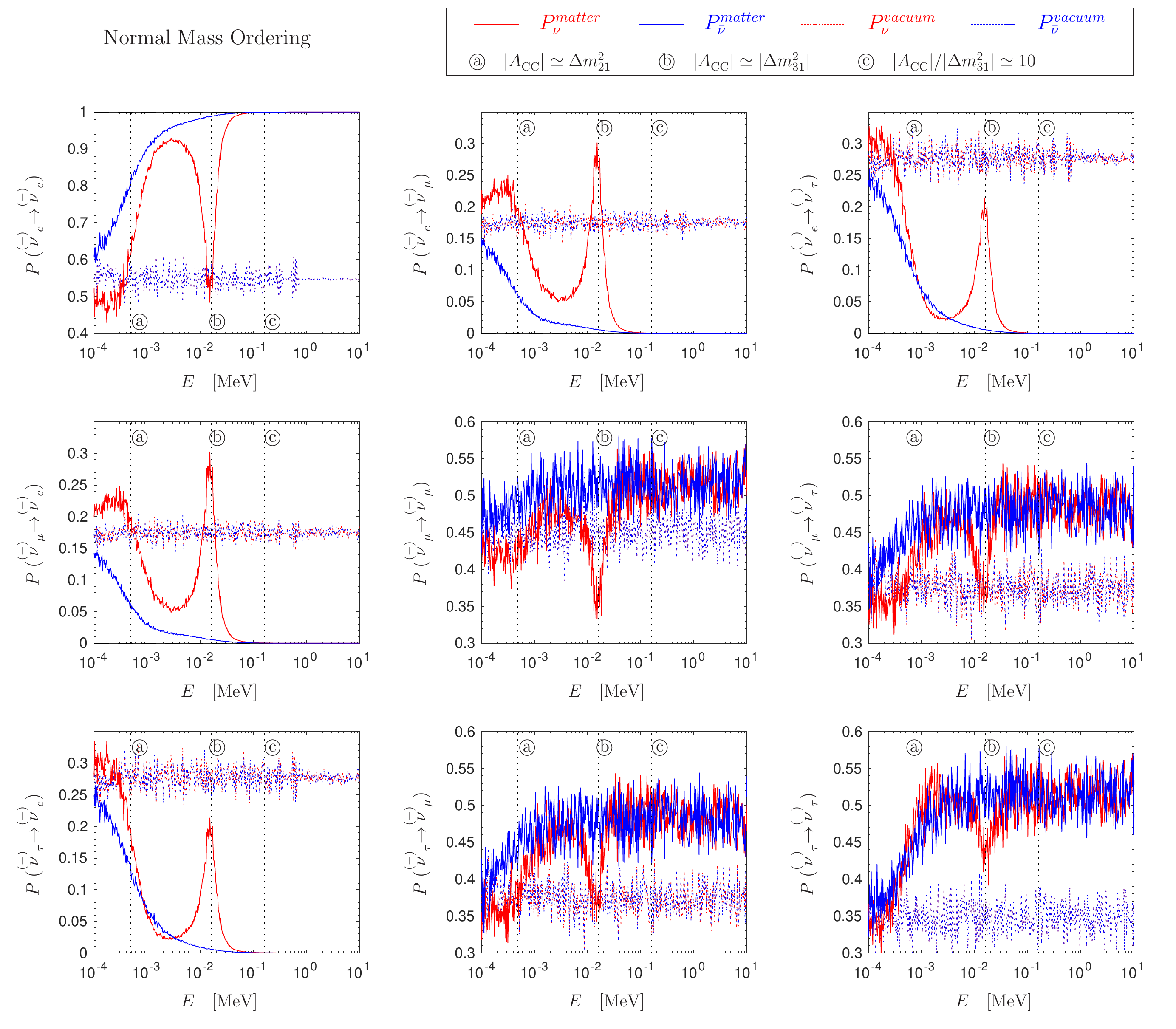}
\vspace{0cm}
\caption{The comparison of the neutrino (anti-neutrino) oscillation probabilities with or without the matter effect as a neutrino (anti-neutrino) beam of energy $E$ go through a typical white dwarf along its diameter, where the normal neutrino mass ordering is assumed and the best-fit values of the mass-squared differences and the mixing parameters in Table. I have been input. The white dwarf is assumed to have an approximately constant density of $\rho \simeq2 \times 10^6 ~ {\rm g}\cdot{\rm cm}^{-3}$ (or equivalently a electron number density of $n^{}_{e} \simeq 10^6 \; N^{}_{\rm A} ~ {\rm cm}^{-3}$ with $N^{}_{\rm A}$ being the Avogadro’s number) and a radius of $R \simeq 10^4$ km. The fixed points of the probabilities in the limit $|A^{}_{\rm CC}| \gg |\Delta m^{2}_{31}|$ given by Eq. (9) (dashed lines) are also plotted in this figure for comparison. Note that all the probabilities are averaged over a Gaussian energy resolution of $5\%$.}
\end{center}
\end{figure}

\begin{figure}[!h]
\begin{center}
\vspace{0cm}
\includegraphics[width=\textwidth]{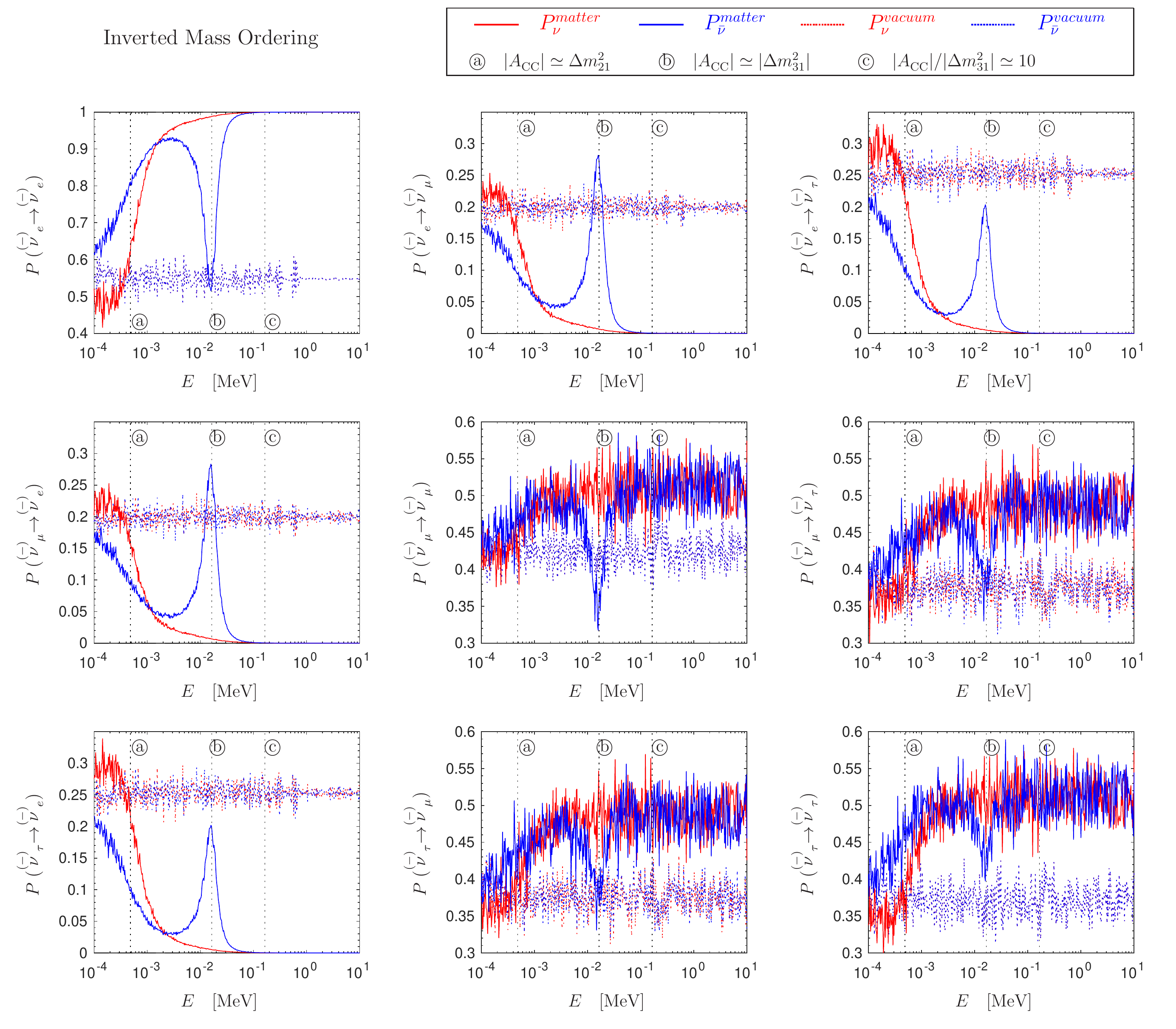}
\vspace{0cm}
\caption{The comparison of the neutrino (anti-neutrino) oscillation probabilities with or without the matter effect as a neutrino (anti-neutrino) beam of energy $E$ go through a typical white dwarf along its diameter, where the inverted neutrino mass ordering is assumed and the best-fit values of the mass-squared differences and the mixing parameters in Table. I have been input. The white dwarf is assumed to have an approximately constant density of $\rho \simeq2 \times 10^6 ~ {\rm g}\cdot{\rm cm}^{-3}$ (or equivalently a electron number density of $n^{}_{e} \simeq 10^6 \; N^{}_{\rm A} ~ {\rm cm}^{-3}$ with $N^{}_{\rm A}$ being the Avogadro’s number) and a radius of $R \simeq 10^4$ km. The fixed points of the probabilities in the limit $|A^{}_{\rm CC}| \gg |\Delta m^{2}_{31}|$ given by Eq. (9) (dashed lines) are also plotted in this figure for comparison. Note that all the probabilities are averaged over a Gaussian energy resolution of $5\%$.}
\end{center}
\end{figure}

Note that all the interesting features of the probabilities we discussed above will finally be embodied in the neutrino/anti-neutrino spectrum we observed. The finding of a change of the slope (around the solar resonance) and a subsequent hump (around the atmospheric resonance) could help to ping down the corresponding resonance energy which can then be turned into the electron density of the compact object. What's more, if both the neutrino and anti-neutrino spectrum can be measured, the present or the absent of the atmospheric resonance hump would be a novel judgement of the neutrino mass ordering. If at a much higher energy range in the matter-dominated region, the oscillatory behavior between $\nu^{}_{\mu}$ and $\nu^{}_{\tau}$ can be observed, the corresponding oscillation frequency, if well determined, may also reveal some information relating to the size of this compact object.

Since white dwarf is a high-density object, there is a concern about the absorption of neutrinos/anti-neutrinos inside the white dwarf. We give a quick estimate of neutrino's mean free path in a typical white dwarf here to preliminarily discuss the significance of this effect for neutrinos with different initial energies. The absorption of neutrinos inside the white dwarf is dominated by the charged-current interaction between neutrinos and the nucleons in the medium. Without loss of generality, we simply use the $\nu$-$n$ (or $\bar{\nu}$-$p$) cross section to evaluate this interaction rate which in the low energy region can be approximately calculated by $\sigma^{\nu n, \bar{\nu} p}_{\rm CC} \simeq 9.3 \times 10^{-44} \left ( E / {\rm MeV} \right )^{-2} {\rm cm}^2$ (see e.g., \cite{Giunti:2007ry,Xing:2011zza}). Then the corresponding mean free path of neutrinos/anti-neutrinos can be written as $\ell = (\sigma \rho / m^{}_{p})^{-1} \sim 0.9 \times 10^{13} \left ( E / {\rm MeV} \right )^{-2} {\rm cm}$, where the typical density of the white dwarf $\rho \sim 2 \times 10^{6} ~ {\rm g} / {\rm cm}^{3}$ have been taken into account. We can then infer from this result, for neutrinos with energy $E \lesssim 10 ~ {\rm MeV}$, the mean free path $\ell \gtrsim 9 \times 10^5 ~ {\rm km}$ can be obtained, which is much larger than the length $2 R \sim 2 \times 10^4 ~ {\rm km}$ the neutrinos transport in the white dwarf. Or in other words, for the neutrino energy of interest to us ($E \lesssim 10 ~ {\rm MeV}$), the white dwarf can be approximately regarded as transparent. Of course if neutrinos with energy higher than $10 ~ {\rm MeV}$ are considered, the attenuation of neutrinos/anti-neutrinos due to both the absorption and scattering need to be carefully studied.

Truly, we cannot actually conduct a long baseline neutrino oscillation experiment on a white dwarf. However, we are now observing neutrinos with a broad range of energies from distant objects using varieties of neutrino detectors, many of which cover the MeV range. If there happen to be a compact object sitting in between the source and the observer, this compact object can not only bend the light and produce the gravitational lensing effect, but also ``lens'' the neutrinos from the source by distorting its spectrum. But different from the gravitational lensing effect which is capable of uncovering the mass distribution in our universe, this ``neutrino lensing'' effect could be sensitive to the distribution of electrons (or positrons) in the space.

Of course, the discussion so far is just an immature and inaccurate thought. For illustrative purposes, the examples we introduced in this manuscript are very simplified and idealized. Lots of details such as the spectrum and the flavor composition of the neutrino source, the properties of the compact objects and their distribution in the space, the capability of the detector have to be carefully studied before we can finally draw the conclusion if this kind of ``neutrino lensing'' effect can be actually observed. In our opinion, it is worthwhile to concentrate more efforts on this topic, for it may open a new window to the universe via the weak interaction of neutrino with the compact objects. We believe that with the improving of the detector capabilities and the data analysis techniques, it is possible to site experiments some day to located the hidden compact objects in the space via this ``neutrino lensing'' effect.

\begin{acknowledgements}

The author is grateful to Prof. Zhi-zhong Xing for bringing up the idea of studying the matter effect in the matter-dominated case and for the enlightening discussions. 
This work is supported by the National Natural Science Foundation of China under Grant No. 11775183.

\end{acknowledgements}

\appendix

\section{Diagonalization of the effective Hamiltonian using the perturbation theory and the fixed points in the limit $\bm{|A^{}_{\rm CC}| \rightarrow \infty}$}

The effective Hamiltonian $\tilde{\cal{H}}$ in the flavor basis responsible for the propagation of neutrinos in matter can be written as 
\begin{eqnarray}
\tilde{\cal{H}} & = & \frac{1}{2 E} \left [ m^{2}_{1} \cdot \mathbbm{1} + V \left ( \begin{matrix} ~ 0 ~ & & \cr & \Delta m^{2}_{21} & \cr & & \Delta m^{2}_{31} \cr \end{matrix} \right ) V^{\dagger}_{} + \left ( \begin{matrix} A^{}_{\rm CC} & & \cr & ~ 0 ~ & \cr & & ~ 0 ~ \cr \end{matrix} \right ) +  A^{}_{\rm NC} \cdot \mathbbm{1} \right ] \; ,
\end{eqnarray}
where  $A^{}_{\rm CC} = 2 E V^{}_{\rm CC}$, $A^{}_{\rm NC} = 2 E V^{}_{\rm NC}$ with $V^{}_{\rm CC} = \sqrt{2} G^{}_{\rm F} N^{}_{e}$ and $\displaystyle V^{}_{\rm NC} = - \frac{\sqrt{2}}{2} G^{}_{\rm F} N^{}_{n}$ being the effective matter potentials describing the charged- and neutral-current interactions of neutrinos with the background $e$, $p$ and $n$ in the medium. Here $G^{}_{\rm F}$ is the Fermi constant, $N^{}_{e}$ and $N^{}_{n}$ are the electron and neutron number densities of the medium respectively, and $\mathbbm{1}$ stands for the $3 \times 3$ unit matrix. Note that the contribution of $\displaystyle \frac{1}{2 E} \left (  m^{2}_{1} + A^{}_{\rm NC} \right ) \cdot \mathbbm{1}$ are identical for all three flavors and therefore would not affect the neutrino oscillation behaviors in matter. In this manuscript we use letters with tilde marks ($\tilde{~}$) to denote the effective neutrino mass and mixing parameters in matter, and letters without the $\tilde{~}$ stand for corresponding parameters in vacuum. The effective Hamiltonian $\tilde{\cal{H}}$ can be diagonalized through a unitary transformation $\tilde{\cal{H}} = \tilde{V} \tilde{\Lambda} \tilde{V}^{\dagger}_{}$, where $ \displaystyle \tilde{\Lambda} = {\rm diag} \{ \tilde{\lambda}^{}_{1}, \tilde{\lambda}^{}_{2}, \tilde{\lambda}^{}_{3} \} = \frac{1}{2 E} {\rm diag} \{ \tilde{m}^{2}_{1}, \tilde{m}^{2}_{2}, \tilde{m}^{2}_{3} \}$ is diagonal with $\tilde{m}^{}_{i} ~ (i = 1, 2, 3)$ being the effective neutrino masses in matter, and the unitary matrix $\tilde{V} = \left ( \tilde{v}^{}_{1} \; \tilde{v}^{}_{2} \; \tilde{v}^{}_{3} \right )$ is just the effective mixing matrix in matter.

In the case of neutrinos having extremely high energy or going through extremely dense object, we could have $|A^{}_{\rm CC}| \gg |\Delta m^{2}_{31}|$ which indicates that the matter potential terms dominate over the vacuum terms. In this matter-dominated region, we may regard both $|\Delta m^{2}_{31} / A^{}_{\rm CC}|$ and $\Delta m^{2}_{21} / |A^{}_{\rm CC}|$ as small parameters and perform the diagonalization of $\tilde{\cal{H}}$ using the perturbation theory. We can then write down the series expansion of the effective Hamiltonian $\tilde{\cal{H}}$ as 
\begin{eqnarray}
\tilde{\cal{H}} & = & \tilde{\cal{H}}^{(0)}_{} + \tilde{\cal{H}}^{(1)}_{} \; ,
\end{eqnarray}
with
\begin{eqnarray}
\tilde{\cal{H}}^{(0)}_{} & = & \frac{1}{2 E} \left [ \left ( m^{2}_{1} + A^{}_{\rm NC} \right ) \cdot \mathbbm{1} + \left ( \begin{matrix} A^{}_{\rm CC} & & \cr & ~ 0 ~ & \cr & & ~ 0 ~ \cr \end{matrix} \right ) \right ]  \; , \\[2mm]
\tilde{\cal{H}}^{(1)}_{} & = & \frac{1}{2 E} \; V \left ( \begin{matrix} ~ 0 ~ & & \cr & \Delta m^{2}_{21} & \cr & & \Delta m^{2}_{31} \cr \end{matrix} \right ) V^{\dagger}_{} \; .
\end{eqnarray}
The eigenvalues and eigenvectors can also be written as $\tilde{\lambda}^{}_{i} = \tilde{\lambda}^{(0)}_{i} + \tilde{\lambda}^{(0)}_{i} +...$ and $\tilde{v}^{}_{i} = \tilde{v}^{(0)}_{i} + \tilde{v}^{(0)}_{i} + ...$ (for $i = 1, 2, 3$) correspondingly. One may immediately find that the zeroth order Hamiltonian $\tilde{\cal{H}}^{(0)}_{}$ is diagonal by itself in the flavor basis, which means
\begin{eqnarray}
\tilde{\lambda}^{(0)}_{1} & = & \frac{1}{2 E} \left ( m^{2}_{1} + A^{}_{\rm NC} + A^{}_{\rm CC} \right ) \; , \nonumber\\[1mm]
\tilde{\lambda}^{(0)}_{2} & = & \frac{1}{2 E} \left ( m^{2}_{1} + A^{}_{\rm NC} \right ) \; , \nonumber\\[1mm]
\tilde{\lambda}^{(0)}_{3} & = & \frac{1}{2 E} \left ( m^{2}_{1} + A^{}_{\rm NC} \right ) \; .
\end{eqnarray}
Note that two eigenvalues of $\tilde{\cal{H}}^{(0)}_{}$ ($\tilde{\lambda}^{(0)}_{2}$ and $\tilde{\lambda}^{(0)}_{3}$) are identical (degenerate). In this case the corresponding zeroth order mixing matrix $\tilde{V}^{(0)}_{}$ should be written as
\begin{eqnarray}
\tilde{V}^{(0)}_{} & = & \left ( \begin{matrix} ~ 1 ~ & ~ 0 ~ & ~ 0 ~ \cr ~ 0 ~ & \cos \tilde{\theta} & \sin \tilde{\theta} \; e^{i \tilde{\phi}}_{} \cr ~ 0 ~ & - \sin \tilde{\theta} \; e^{-i \tilde{\phi}}_{} & \cos \tilde{\theta} \cr \end{matrix} \right )  \; .
\end{eqnarray}
By carefully repeating the derivation, we find that, in the case of $\tilde{\cal{H}}^{(0)}_{}$ possessing two degenerate eigenvalues (e.g., $\tilde{\lambda}^{(0)}_{2} = \tilde{\lambda}^{(0)}_{3}$), if $ <\tilde{\cal{H}}>^{(n)}_{23} = <\tilde{\cal{H}}>^{(n)}_{32} = 0$ are satisfied for any integer $n \ge 1$, one could still solving three eigenvalues $\tilde{\lambda}^{}_{i}$ and the unitarity transformational matrix $\tilde{V}$ using the perturbation theory, where $<\tilde{\cal{H}}>^{(n)}_{ij} \equiv \tilde{v}^{(0) \dagger}_{i} \tilde{\cal{H}}^{(n)}_{} \tilde{v}^{(0)}_{j}$ (i.e., $<\tilde{\cal{H}}>^{(n)}_{} \equiv \tilde{V}^{(0) \dagger}_{} \tilde{\cal{H}}^{(n)}_{} \tilde{V}^{(0)}_{}$). Above conditions are obviously satisfied for any $n \ge 2$, since we have $\tilde{\cal{H}}^{(n)}_{} = 0$ (for $n \ge 2$) as one can find from Eq. (A2). And further more, from $<\tilde{\cal{H}}>^{(1)}_{23} = <\tilde{\cal{H}}>^{(1)}_{32} = 0$, it's quite straightforward to have $\tilde{\theta}$ and $\tilde{\phi}$ solved as
\begin{eqnarray}
\tan 2 \tilde{\theta} & = & \frac{2 | \Omega^{}_{23} |}{\Omega^{}_{33} - \Omega^{}_{22}} \; = \; \frac{2 | \Delta m^{2}_{21} V^{}_{\mu 2} V^{*}_{\tau 2}  + \Delta m^{2}_{31} V^{}_{\mu 3} V^{*}_{\tau 3} |}{\Delta m^{2}_{21} \left ( | V^{}_{\tau 2} |^{2}_{} - | V^{}_{\mu 2} |^{2}_{} \right ) + \Delta m^{2}_{31} \left ( | V^{}_{\tau 3} |^{2}_{} - | V^{}_{\mu 3} |^{2}_{} \right )} \; , \nonumber\\[2mm]
\tilde{\phi} & = & \arg \left ( \Omega^{}_{23} \right ) \; = \; \arg \left ( \Delta m^{2}_{21} V^{}_{\mu 2} V^{*}_{\tau 2}  + \Delta m^{2}_{31} V^{}_{\mu 3} V^{*}_{\tau 3} \right ) \; .
\end{eqnarray}
Here the Hermitian matrix $\Omega$ is defined as
\begin{eqnarray}
\Omega & \equiv & V \left ( \begin{matrix} ~ 0 ~ & & \cr & \Delta m^{2}_{21} & \cr & & \Delta m^{2}_{31} \cr \end{matrix} \right ) V^{\dagger}_{} \nonumber\\
& = & \left ( \begin{matrix} \Delta m^{2}_{21} |V^{}_{e 2}|^{2}_{} + \Delta m^{2}_{31} |V^{}_{e 3}|^{2}_{} & \Delta m^{2}_{21} V^{}_{e 2} V^{*}_{\mu 2} + \Delta m^{2}_{31} V^{}_{e 3} V^{*}_{\mu 3} & \Delta m^{2}_{21} V^{}_{e 2} V^{*}_{\tau 2} + \Delta m^{2}_{31} V^{}_{e 3} V^{*}_{\tau 3} \cr \Delta m^{2}_{21} V^{}_{\mu 2} V^{*}_{e 2} + \Delta m^{2}_{31} V^{}_{\mu 3} V^{*}_{e 3} & \Delta m^{2}_{21} |V^{}_{\mu 2}|^{2}_{} + \Delta m^{2}_{31} |V^{}_{\mu 3}|^{2}_{} & \Delta m^{2}_{21} V^{}_{\mu 2} V^{*}_{\tau 2} + \Delta m^{2}_{31} V^{}_{\mu 3} V^{*}_{\tau 3} \cr \Delta m^{2}_{21} V^{}_{\tau 2} V^{*}_{e 2} + \Delta m^{2}_{31} V^{}_{\tau 3} V^{*}_{e 3} & \Delta m^{2}_{21} V^{}_{\tau 2} V^{*}_{\mu 2} + \Delta m^{2}_{31} V^{}_{\tau 3} V^{*}_{\mu 3} & \Delta m^{2}_{21} |V^{}_{\tau 2}|^{2}_{} + \Delta m^{2}_{31} |V^{}_{\tau 3}|^{2}_{} \cr \end{matrix} \right ) \;. \nonumber\\
\end{eqnarray}

Given that the zeroth order solutions are well determined, the first order corrections to the eigenvalues and eigenvectors can be expressed as
\begin{eqnarray}
\tilde{\lambda}^{(1)}_{1} & = & < \tilde{\cal H} >^{(1)}_{11} \; = \; \frac{\Omega^{}_{11}}{2 E} \; , \nonumber\\[2mm]
\tilde{\lambda}^{(1)}_{2} & = & < \tilde{\cal H} >^{(1)}_{22} \; = \; \frac{1}{2 E} \left ( \Omega^{}_{22} \cos^2\tilde{\theta} +\Omega^{}_{33} \sin^2\tilde{\theta} - | \Omega^{}_{23} | \sin2\tilde{\theta} \right ) \; , \nonumber\\[2mm]
\tilde{\lambda}^{(1)}_{3} & = & < \tilde{\cal H} >^{(1)}_{33} \; = \; \frac{1}{2 E} \left ( \Omega^{}_{33} \cos^2\tilde{\theta} +\Omega^{}_{22} \sin^2\tilde{\theta} + | \Omega^{}_{23} | \sin2\tilde{\theta} \right ) \; ,
\end{eqnarray}
and
\begin{eqnarray}
\tilde{v}^{(1)}_{1} & = & \frac{< \tilde{\cal H} >^{(1)}_{21}}{\tilde{\lambda}^{(0)}_{1} - \tilde{\lambda}^{(0)}_{2}} \cdot \tilde{v}^{(0)}_{2} + \frac{< \tilde{\cal H} >^{(1)}_{31}}{\tilde{\lambda}^{(0)}_{1} - \tilde{\lambda}^{(0)}_{3}} \cdot \tilde{v}^{(0)}_{3} \; = \; \frac{1}{A^{}_{\rm CC}} \left ( \begin{matrix} 0 \cr \Omega^{}_{21} \cr \Omega^{}_{31} \cr \end{matrix} \right ) \; , \nonumber\\[2mm]
\tilde{v}^{(1)}_{2} & = & \frac{< \tilde{\cal H} >^{(1)}_{12}}{\tilde{\lambda}^{(0)}_{2} - \tilde{\lambda}^{(0)}_{1}} \cdot \tilde{v}^{(0)}_{1} \; = \; \frac{1}{A^{}_{\rm CC}} \left ( \begin{matrix} - \Omega^{}_{12} \cos\tilde{\theta} + \Omega^{}_{13} \sin\tilde{\theta} \; e^{-i \tilde{\phi}}_{} \cr 0 \cr 0 \cr \end{matrix} \right ) \; , \nonumber\\[2mm]
\tilde{v}^{(1)}_{3} & = & \frac{< \tilde{\cal H} >^{(1)}_{13}}{\tilde{\lambda}^{(0)}_{3} - \tilde{\lambda}^{(0)}_{1}} \cdot \tilde{v}^{(0)}_{1} \; = \; \frac{1}{A^{}_{\rm CC}} \left ( \begin{matrix} - \Omega^{}_{13} \cos\tilde{\theta} - \Omega^{}_{12} \sin\tilde{\theta} \; e^{i \tilde{\phi}}_{} \cr 0 \cr 0 \cr \end{matrix} \right ) \; .
\end{eqnarray}
One can clearly see that the lowest order corrections to $\tilde{V}^{(0)}_{e 1}$, $\tilde{V}^{(0)}_{\mu 2}$, $\tilde{V}^{(0)}_{\mu 3}$, $\tilde{V}^{(0)}_{\tau 2}$ and $\tilde{V}^{(0)}_{\tau 3}$ come only in the second order corrections which are highly suppressed if $A^{}_{\rm CC}$ dominates. And the second order corrections to the eigenvalues are given by
\begin{eqnarray}
\tilde{\lambda}^{(2)}_{1} & = & \frac{| < \tilde{\cal H} >^{(1)}_{21} |^{2}_{}}{\tilde{\lambda}^{(0)}_{1} - \tilde{\lambda}^{(0)}_{2}} + \frac{| < \tilde{\cal H} >^{(1)}_{31} |^{2}_{}}{\tilde{\lambda}^{(0)}_{1} - \tilde{\lambda}^{(0)}_{3}} \; = \; \frac{1}{2 E A^{}_{\rm CC}} \left ( | \Omega^{}_{12} |^{2}_{} + | \Omega^{}_{13} |^{2}_{} \right )  \; , \nonumber\\[2mm]
\tilde{\lambda}^{(2)}_{2} & = & \frac{| < \tilde{\cal H} >^{(1)}_{12} |^{2}_{}}{\tilde{\lambda}^{(0)}_{2} - \tilde{\lambda}^{(0)}_{1}} \; = \; - \frac{1}{2 E A^{}_{\rm CC}} \left | \Omega^{}_{12} \cos\tilde{\theta} - \Omega^{}_{13} \sin\tilde{\theta} e^{- i \tilde{\phi}}_{} \right |^{2}_{} \; , \nonumber\\[2mm]
\tilde{\lambda}^{(2)}_{3} & = & \frac{| < \tilde{\cal H} >^{(1)}_{13} |^{2}_{}}{\tilde{\lambda}^{(0)}_{3} - \tilde{\lambda}^{(0)}_{1}} \; = \; - \frac{1}{2 E A^{}_{\rm CC}} \left | \Omega^{}_{13} \cos\tilde{\theta} + \Omega^{}_{12} \sin\tilde{\theta} e^{i \tilde{\phi}}_{} \right |^{2}_{} \; .
\end{eqnarray}
Note that, the effective Hamiltonian $\tilde{\cal{H}}$ itself contains terms proportional to $A^{}_{\rm CC}$ thus the two differences $\tilde{\lambda}^{(0)}_{1} - \tilde{\lambda}^{(0)}_{2}$, $\tilde{\lambda}^{(0)}_{1} - \tilde{\lambda}^{(0)}_{3}$ are also proportional to $A^{}_{\rm CC}$. As a result, the expansions of the eigenvalues $\tilde{\lambda}^{(n)}_{i}$ (for $i = 1, 2, 3$) are actually of the order ${\cal O} (\Delta m^{2}_{ji} / A^{}_{\rm CC})^{n-1}_{}$ (for $ji = 31$ or $21$), while  the expansions of the eigenvectors $\tilde{v}^{(n)}_{i}$ (for $i = 1, 2, 3$) are still of the order ${\cal O} (\Delta m^{2}_{ji} / A^{}_{\rm CC})^{n}_{}$ (for $ji = 31$ or $21$).

To sum up, up to the first order of both $|\Delta m^{2}_{31} / A^{}_{\rm CC}|$ and $\Delta m^{2}_{21} / |A^{}_{\rm CC}|$, three eigenvalues of $\tilde{\cal H}$ and the effective mixing matrix in matter can be approximately expressed as 
{\small
\begin{eqnarray}
\tilde{\Lambda} 
& \approx & \frac{1}{2 E} \left ( m^{2}_{1} + A^{}_{\rm NC} \right ) \cdot \mathbbm{1} \nonumber\\[2mm]
& & + \frac{1}{2 E} \left ( \begin{matrix} A^{}_{\rm CC} + \Omega^{}_{11} & & \cr & \Omega^{}_{22} \cos^2\tilde{\theta} +\Omega^{}_{33} \sin^2\tilde{\theta} - | \Omega^{}_{23} | \sin2\tilde{\theta} & \cr & & \Omega^{}_{33} \cos^2\tilde{\theta} +\Omega^{}_{22} \sin^2\tilde{\theta} + | \Omega^{}_{23} | \sin2\tilde{\theta} \cr \end{matrix} \right ) \nonumber\\[2mm]
& & + \frac{1}{2 E A^{}_{\rm CC}} \left ( \begin{matrix}  | \Omega^{}_{12} |^{2}_{} + | \Omega^{}_{13} |^{2}_{} & & \cr &- \left | \Omega^{}_{12} \cos\tilde{\theta} - \Omega^{}_{13} \sin\tilde{\theta} e^{- i \tilde{\phi}}_{} \right |^{2}_{} & \cr & & - \left | \Omega^{}_{13} \cos\tilde{\theta} + \Omega^{}_{12} \sin\tilde{\theta} e^{i \tilde{\phi}}_{} \right |^{2}_{} \cr \end{matrix} \right ) \; ,  \nonumber\\
\end{eqnarray}}
and
\begin{eqnarray}
\tilde{V} & \approx & \left ( \begin{matrix} 1 & \displaystyle ~ \frac{- \Omega^{}_{12} \cos\tilde{\theta} + \Omega^{}_{13} \sin\tilde{\theta} \; e^{-i \tilde{\phi}}_{}}{A^{}_{\rm CC}} ~ & \displaystyle ~ \frac{- \Omega^{}_{13} \cos\tilde{\theta} - \Omega^{}_{12} \sin\tilde{\theta} \; e^{i \tilde{\phi}}_{}}{A^{}_{\rm CC}} ~ \\[3mm] \displaystyle ~ \frac{\Omega^{}_{21}}{A^{}_{\rm CC}} ~ & \cos \tilde{\theta} & \sin \tilde{\theta} \; e^{i \tilde{\phi}}_{} \\[3mm] \displaystyle ~ \frac{\Omega^{}_{31}}{A^{}_{\rm CC}} ~ & - \sin \tilde{\theta} \; e^{-i \tilde{\phi}}_{} & \cos \tilde{\theta} \cr \end{matrix} \right ) \; .
\end{eqnarray}

If the matter density can be regarded as a constant along the path neutrinos propagate, we can then write down the neutrino oscillation probabilities in matter simply by replacing the neutrino mass-squared differences and the mixing matrix in neutrino oscillation probabilities in vacuum with the corresponding effective neutrino mass and mixing parameters in matter. 
\begin{eqnarray}
\tilde{P} \; ( \stackrel{(-)}{\nu}^{}_{\alpha} \rightarrow  \stackrel{(-)}{\nu}^{}_{\beta} ) & = & \delta^{}_{\alpha \beta} - 4 \sum^{}_{j > i} {\rm Re} \left [ \tilde{V}^{}_{\alpha i} \tilde{V}^{}_{\beta j} \tilde{V}^{*}_{\alpha j} \tilde{V}^{*}_{\beta i} \right ] \sin^2 \tilde{\Delta}^{}_{ji} \pm 2 \sum^{}_{j > i} {\rm Im} \left [ \tilde{V}^{}_{\alpha i} \tilde{V}^{}_{\beta j} \tilde{V}^{*}_{\alpha j} \tilde{V}^{*}_{\beta i} \right ] \sin2\tilde{\Delta}^{}_{ji} \; , ~~~~~~~
\end{eqnarray}
where $\tilde{\Delta}^{}_{ji} \equiv \Delta \tilde{m}^2_{ji} L / 4 E$ with $\Delta \tilde{m}^2_{ji} \equiv \tilde{m}^{2}_{j} - \tilde{m}^{2}_{i} = 2 E ( \tilde{\lambda}^{}_{j} - \tilde{\lambda}^{}_{i} )$ being the effective neutrino mass-squared difference in matter. Here the Greek letters $\alpha$, $\beta$ are flavor indices run over $e$, $\mu$, $\tau$, while the Latin letters $i$, $j$ are indices of the mass eigenstates run over $1$, $2$, $3$. Using the results summarized in Eqs. (A12) and (A13), the neutrino oscillation probabilities in matter to the second order of both $|\Delta m^{2}_{31} / A^{}_{\rm CC}|$ and $\Delta m^{2}_{21} / |A^{}_{\rm CC}|$ can be expressed as
\begin{eqnarray}
\tilde{P}(\nu^{}_{e} \rightarrow \nu^{}_{e}) & \approx & 1 - \frac{4}{A^{2}_{\rm CC}} \left | \Omega^{}_{12} \cos\tilde{\theta} - \Omega^{}_{13} \sin\tilde{\theta} e^{- i \tilde{\phi}}_{} \right |^{2}_{} \sin^2 \frac{\Delta \tilde{m}^{2}_{21} L}{4 E} \nonumber\\[1mm]
&& ~ - \frac{4}{A^{2}_{\rm CC}} \left | \Omega^{}_{13} \cos\tilde{\theta} + \Omega^{}_{12} \sin\tilde{\theta} e^{i \tilde{\phi}}_{} \right |^{2}_{} \sin^2 \frac{\Delta \tilde{m}^{2}_{31} L}{4 E} \; , \nonumber\\[2mm]
\tilde{P}(\nu^{}_{\mu} \rightarrow \nu^{}_{\mu}) &\approx & 1 - \sin^2 2\tilde{\theta} \sin^2 \frac{\Delta \tilde{m}^{2}_{32} L}{4 E} \nonumber\\[1mm]
&& ~ - \frac{4}{A^{2}_{\rm CC}} | \Omega^{}_{12} |^{2}_{} \left ( \cos^2\tilde{\theta} \sin^2 \frac{\Delta \tilde{m}^{2}_{21} L}{4 E} + \sin^2\tilde{\theta} \sin^2 \frac{\Delta \tilde{m}^{2}_{31} L}{4 E} \right ) \; , \nonumber\\[2mm]
\tilde{P}(\nu^{}_{\tau} \rightarrow \nu^{}_{\tau}) & \approx & 1 - \sin^2 2\tilde{\theta} \sin^2 \frac{\Delta \tilde{m}^{2}_{32} L}{4 E} \nonumber\\[1mm]
&& ~ - \frac{4}{A^{2}_{\rm CC}} | \Omega^{}_{13} |^{2}_{} \left ( \cos^2\tilde{\theta} \sin^2 \frac{\Delta \tilde{m}^{2}_{21} L}{4 E} + \sin^2\tilde{\theta} \sin^2 \frac{\Delta \tilde{m}^{2}_{31} L}{4 E} \right ) \; , \nonumber\\[2mm]
\tilde{P}(\nu^{}_{e} \rightarrow \nu^{}_{\mu}) & \approx & \frac{1}{A^{2}_{\rm CC}} \left \{ \left ( 4 | \Omega^{}_{12} |^{2}_{} \cos^2\tilde{\theta} - 2 {\rm Re} [\Omega^{*}_{13} \Omega^{}_{12} e^{i \tilde{\phi}}_{}] \sin2\tilde{\theta} \right ) \sin^2 \frac{\Delta \tilde{m}^{2}_{21} L}{4 E} \right. \nonumber\\[1mm]
&& ~~~~~~ + \left ( 4 | \Omega^{}_{12} |^{2}_{} \sin^2\tilde{\theta} + 2 {\rm Re} [\Omega^{*}_{13} \Omega^{}_{12} e^{i \tilde{\phi}}_{}]  \sin2\tilde{\theta} \right ) \sin^2 \frac{\Delta \tilde{m}^{2}_{31} L}{4 E} \nonumber\\[1mm]
&& ~~~~~~ + \left ( ( | \Omega^{}_{13} |^{2}_{} - | \Omega^{}_{12} |^{2}_{} ) \sin^2 2\tilde{\theta} - {\rm Re} [\Omega^{*}_{13} \Omega^{}_{12} e^{i \tilde{\phi}}_{}] \sin4\tilde{\theta} \right ) \sin^2 \frac{\Delta \tilde{m}^{2}_{32} L}{4 E} \nonumber\\[1mm]
&& ~~~~~~ \left . + {\rm Im} [\Omega^{*}_{13} \Omega^{}_{12} e^{i \tilde{\phi}}_{}] \sin2\tilde{\theta} \left ( \sin \frac{\Delta \tilde{m}^{2}_{21} L}{2 E} - \sin \frac{\Delta \tilde{m}^{2}_{31} L}{2 E} + \sin \frac{\Delta \tilde{m}^{2}_{32} L}{2 E} \right ) \right \} \; , \nonumber\\[2mm]
\tilde{P}(\nu^{}_{e} \rightarrow \nu^{}_{\tau}) & \approx & \frac{1}{A^{2}_{\rm CC}} \left \{ \left ( 4 | \Omega^{}_{13} |^{2}_{} \sin^2\tilde{\theta} - 2 {\rm Re} [\Omega^{*}_{13} \Omega^{}_{12} e^{i \tilde{\phi}}_{}] \sin2\tilde{\theta} \right ) \sin^2 \frac{\Delta \tilde{m}^{2}_{21} L}{4 E} \right. \nonumber\\[1mm]
&& ~~~~~~ + \left ( 4 | \Omega^{}_{13} |^{2}_{} \cos^2\tilde{\theta} + 2 {\rm Re} [\Omega^{*}_{13} \Omega^{}_{12} e^{i \tilde{\phi}}_{}] \sin2\tilde{\theta} \right ) \sin^2 \frac{\Delta \tilde{m}^{2}_{31} L}{4 E} \nonumber\\[1mm]
&& ~~~~~~ - \left ( ( | \Omega^{}_{13} |^{2}_{} - | \Omega^{}_{12} |^{2}_{} ) \sin^2 2\tilde{\theta} - {\rm Re} [\Omega^{*}_{13} \Omega^{}_{12} e^{i \tilde{\phi}}_{}] \sin4\tilde{\theta} \right ) \sin^2 \frac{\Delta \tilde{m}^{2}_{32} L}{4 E} \nonumber\\[1mm]
&& ~~~~~~ \left . - {\rm Im} [\Omega^{*}_{13} \Omega^{}_{12} e^{i \tilde{\phi}}_{}] \sin2\tilde{\theta} \left ( \sin \frac{\Delta \tilde{m}^{2}_{21} L}{2 E} - \sin \frac{\Delta \tilde{m}^{2}_{31} L}{2 E} + \sin \frac{\Delta \tilde{m}^{2}_{32} L}{2 E} \right ) \right \} \; , \nonumber\\[2mm]
\tilde{P}(\nu^{}_{\mu} \rightarrow \nu^{}_{\tau}) & \approx & \sin^2 2\tilde{\theta} \sin^2 \frac{\Delta \tilde{m}^{2}_{32} L}{4 E} \nonumber\\[1mm]
&& + \frac{1}{A^{2}_{\rm CC}} \left \{ 2 {\rm Re} [\Omega^{*}_{13} \Omega^{}_{12} e^{i \tilde{\phi}}_{}] \sin2\tilde{\theta} \left ( \sin^2 \frac{\Delta \tilde{m}^{2}_{21} L}{4 E} - \sin^2 \frac{\Delta \tilde{m}^{2}_{31} L}{4 E} \right ) \right. \nonumber\\[1mm]
&& ~~~~~~~~~ \left. + {\rm Im} [\Omega^{*}_{13} \Omega^{}_{12} e^{i \tilde{\phi}}_{}] \sin2\tilde{\theta} \left ( \sin \frac{\Delta \tilde{m}^{2}_{21} L}{2 E} - \sin \frac{\Delta \tilde{m}^{2}_{31} L}{2 E} + \sin \frac{\Delta \tilde{m}^{2}_{32} L}{2 E} \right ) \right \} \; , \nonumber\\
\end{eqnarray}
with
\begin{eqnarray}
\Delta \tilde{m}^{2}_{21} & \approx & - A^{}_{\rm CC} - \Omega^{}_{11} + \Omega^{}_{22} \cos^2\tilde{\theta} + \Omega^{}_{33} \sin^2\tilde{\theta} - | \Omega^{}_{23} | \sin2\tilde{\theta} \nonumber\\[1mm]
& & - \frac{1}{A^{}_{\rm CC}} \left(  | \Omega^{}_{12} |^{2}_{} + | \Omega^{}_{13} |^{2}_{} + \left | \Omega^{}_{12} \cos\tilde{\theta} - \Omega^{}_{13} \sin\tilde{\theta} e^{- i \tilde{\phi}}_{} \right |^{2}_{} \right ) \; , \nonumber\\[2mm]
\Delta \tilde{m}^{2}_{31} & \approx & - A^{}_{\rm CC} - \Omega^{}_{11} + \Omega^{}_{33} \cos^2\tilde{\theta} + \Omega^{}_{22} \sin^2\tilde{\theta} + | \Omega^{}_{23} | \sin2\tilde{\theta} \nonumber\\[1mm]
& & - \frac{1}{A^{}_{\rm CC}} \left(  | \Omega^{}_{12} |^{2}_{} + | \Omega^{}_{13} |^{2}_{} + \left | \Omega^{}_{13} \cos\tilde{\theta} + \Omega^{}_{12} \sin\tilde{\theta} e^{i \tilde{\phi}}_{} \right |^{2}_{} \right ) \; , \nonumber\\[2mm]
\Delta \tilde{m}^{2}_{32} & \approx & \left ( \Omega^{}_{33} - \Omega^{}_{22} \right ) \cos2\tilde{\theta} + 2 | \Omega^{}_{23} | \sin2\tilde{\theta} \nonumber\\[1mm]
& & + \frac{1}{A^{}_{\rm CC}} \left( | \Omega^{}_{13} |^{2}_{} - | \Omega^{}_{12} |^{2}_{} ) \cos 2\tilde{\theta} + 2 {\rm Re} [\Omega^{*}_{13} \Omega^{}_{12} e^{i \tilde{\phi}}_{}] \sin2\tilde{\theta} \right ) \; .
\end{eqnarray}
Another three neutrino oscillation probabilities $\tilde{P}(\nu^{}_{\mu} \rightarrow \nu^{}_{e})$, $\tilde{P}(\nu^{}_{\tau} \rightarrow \nu^{}_{e})$ and $\tilde{P}(\nu^{}_{\tau} \rightarrow \nu^{}_{\mu})$ can be obtained by changing the signs of all the ${\rm Im} [\Omega^{*}_{13} \Omega^{}_{12} e^{i \tilde{\phi}}_{}]$ terms in $\tilde{P}(\nu^{}_{e} \rightarrow \nu^{}_{\mu})$, $\tilde{P}(\nu^{}_{e} \rightarrow \nu^{}_{\tau})$ and $\tilde{P}(\nu^{}_{\mu} \rightarrow \nu^{}_{\tau})$ correspondingly.
In addition, one can calculate the anti-neutrino oscillation probabilities in matter $\tilde{P}(\bar{\nu}^{}_{\alpha} \rightarrow \bar{\nu}^{}_{\beta})$ using the following relation
\begin{eqnarray}
\tilde{P}(\bar{\nu}^{}_{\alpha} \rightarrow \bar{\nu}^{}_{\beta}) ( V, A^{}_{\rm CC} ) & = & \tilde{P}(\nu^{}_{\alpha} \rightarrow \nu^{}_{\beta}) ( V^{*}_{}, - A^{}_{\rm CC} )  \; .
\end{eqnarray}
It's worth mentioning again that above formulas are series expansions in both $|\Delta m^{2}_{31} / A^{}_{\rm CC}|$ and $\Delta m^{2}_{21} / |A^{}_{\rm CC}|$, which means they are good approximations only in the region $|A^{}_{\rm CC}| > |\Delta m^{2}_{31}|$, i.e., in the case of neutrinos having extremely high energy or going through extremely dense object.

In the matter-dominated region, as the increase of $|A^{}_{\rm CC}|$,  terms proportional to $1 / A^{}_{\rm CC}$ are all approaching zero fast, and therefore as one can clearly seen from Eqs. (A12) and (A13) that three eigenvalues of $\tilde{\cal H}$ are approaching a set of fixed values
\begin{eqnarray}
\tilde{\lambda}^{fixed}_{1} & \approx & \frac{1}{2 E} \left ( m^{2}_{1} + A^{}_{\rm NC} + A^{}_{\rm CC} + \Omega^{}_{11} \right ) \; , \nonumber\\[1mm]
\tilde{\lambda}^{fixed}_{2} & \approx & \frac{1}{2 E} \left ( m^{2}_{1} + A^{}_{\rm NC} + \Omega^{}_{22} \cos^2\tilde{\theta} +\Omega^{}_{33} \sin^2\tilde{\theta} - | \Omega^{}_{23} | \sin2\tilde{\theta} \right ) \; , \nonumber\\[1mm]
\tilde{\lambda}^{fixed}_{3} & \approx & \frac{1}{2 E} \left ( m^{2}_{1} + A^{}_{\rm NC} + \Omega^{}_{33} \cos^2\tilde{\theta} +\Omega^{}_{22} \sin^2\tilde{\theta} + | \Omega^{}_{23} | \sin2\tilde{\theta} \right ) \; .
\end{eqnarray}
Apparently, in this matter-dominated case, $\tilde{\lambda}^{fixed}_{2}$ and $\tilde{\lambda}^{fixed}_{3}$ are nearly degenerate and both of them have strong hierarchies with $\tilde{\lambda}^{fixed}_{1}$.
In the same time the effective mixing matrix in matter $\tilde{V}$ evolves towards a fixed $3 \times 3$ real matrix
\begin{eqnarray}
\tilde{V}^{fixed}_{} & \approx & \left ( \begin{matrix} ~ 1 ~ & ~ 0 ~ & ~ 0 ~ \cr ~ 0 ~ & \cos \tilde{\theta} & \sin \tilde{\theta} \cr ~ 0 ~ & - \sin \tilde{\theta} & \cos \tilde{\theta} \cr \end{matrix} \right )  \; ,
\end{eqnarray}
which has the two-flavor-mixing structure and can be expressed using just one mixing angle $\tilde{\theta}$ as defined in Eq. (A7). 

In the limit $1 / A^{}_{\rm CC} \rightarrow 0$, those neutrino oscillation probabilities can be concisely expressed as
\begin{eqnarray}
\tilde{P}(\nu^{}_{e} \rightarrow \nu^{}_{e}) & \approx & \tilde{P}(\bar{\nu}^{}_{e} \rightarrow \bar{\nu}^{}_{e}) \; \approx \; 1 \; , \nonumber\\[1mm]
\tilde{P}(\nu^{}_{e} \rightarrow \nu^{}_{\mu}) & \approx & \tilde{P}(\bar{\nu}^{}_{e} \rightarrow \bar{\nu}^{}_{\mu}) \; \approx \; 0 \; , \nonumber\\[1mm]
\tilde{P}(\nu^{}_{e} \rightarrow \nu^{}_{\tau}) & \approx & \tilde{P}(\bar{\nu}^{}_{e} \rightarrow \bar{\nu}^{}_{\tau}) \; \approx \; 0 \; , \nonumber\\[1mm]
\tilde{P}(\nu^{}_{\mu} \rightarrow \nu^{}_{e}) & \approx & \tilde{P}(\bar{\nu}^{}_{\mu} \rightarrow \bar{\nu}^{}_{e}) \; \approx \; 0 \; , \nonumber\\[1mm]
\tilde{P}(\nu^{}_{\mu} \rightarrow \nu^{}_{\mu}) & \approx & \tilde{P}(\bar{\nu}^{}_{\mu} \rightarrow \bar{\nu}^{}_{\mu}) \; \approx \; 1 - \sin^2 2\tilde{\theta} \sin^2 \frac{\Delta \tilde{m}^{2}_{32} L}{4 E} \; , \nonumber\\[1mm]
\tilde{P}(\nu^{}_{\mu} \rightarrow \nu^{}_{\tau}) & \approx & \tilde{P}(\bar{\nu}^{}_{\mu} \rightarrow \bar{\nu}^{}_{\tau}) \; \approx \; \sin^2 2\tilde{\theta} \sin^2 \frac{\Delta \tilde{m}^{2}_{32} L}{4 E} \; , \nonumber\\[1mm]
\tilde{P}(\nu^{}_{\tau} \rightarrow \nu^{}_{e}) & \approx & \tilde{P}(\bar{\nu}^{}_{\tau} \rightarrow \bar{\nu}^{}_{e}) \; \approx \; 0 \; , \nonumber\\[1mm]
\tilde{P}(\nu^{}_{\tau} \rightarrow \nu^{}_{\mu}) & \approx & \tilde{P}(\bar{\nu}^{}_{\tau} \rightarrow \bar{\nu}^{}_{\mu}) \; \approx \; \sin^2 2\tilde{\theta} \sin^2 \frac{\Delta \tilde{m}^{2}_{32} L}{4 E} \; , \nonumber\\[1mm]
\tilde{P}(\nu^{}_{\tau} \rightarrow \nu^{}_{\tau}) & \approx & \tilde{P}(\bar{\nu}^{}_{\tau} \rightarrow \bar{\nu}^{}_{\tau}) \; \approx \; 1 - \sin^2 2\tilde{\theta} \sin^2 \frac{\Delta \tilde{m}^{2}_{32} L}{4 E} \; .
\end{eqnarray}
In this matter-dominated condition, $\nu^{}_{e}$ are decoupled, while oscillations can still happened between $\nu^{}_{\mu}$ and $\nu^{}_{\tau}$. This two-flavor oscillation can be described simply by one effective mixing angle $\tilde{\theta}$ and one effective mass-squared difference $\Delta \tilde{m}^{2}_{32}$ whose expressions are given in Eqs. (A7) and (A16) respectively. One may immediately find that both the parameters are independent of $A^{}_{\rm CC}$ and can be easily calculated once the oscillation parameters in vacuum are well determined.
What's more, in the limit $1 / A^{}_{\rm CC} \rightarrow 0$ we arrive at $\tilde{P}(\bar{\nu}^{}_{\alpha} \rightarrow \bar{\nu}^{}_{\beta}) \approx \tilde{P}(\nu^{}_{\alpha} \rightarrow \nu^{}_{\beta})$, for $\alpha, \beta = e, \mu, \tau$. The intrinsic CP violation asymptotically vanishes in the matter-dominated case.


\begin{thebibliography}{99}


\bibitem{Wolfenstein:1977ue} 
  L.~Wolfenstein,
  Phys.\ Rev.\ D {\bf 17}, 2369 (1978).

\bibitem{Mikheev:1986gs} 
  S.~P.~Mikheyev and A.~Y.~Smirnov,
  Sov.\ J.\ Nucl.\ Phys.\  {\bf 42}, 913 (1985)
  [Yad.\ Fiz.\  {\bf 42}, 1441 (1985)].

\bibitem{Mikheev:1986wj} 
  S.~P.~Mikheev and A.~Y.~Smirnov,
  Nuovo Cim.\ C {\bf 9}, 17 (1986).

\bibitem{Kuo:1989qe} 
  T.~K.~Kuo and J.~T.~Pantaleone,
  Rev.\ Mod.\ Phys.\  {\bf 61}, 937 (1989).



\bibitem{Maki:1962mu} 
  Z.~Maki, M.~Nakagawa and S.~Sakata,
  Prog.\ Theor.\ Phys.\  {\bf 28}, 870 (1962).

\bibitem{Pontecorvo:1967fh} 
  B.~Pontecorvo,
  Sov.\ Phys.\ JETP {\bf 26}, 984 (1968)
  [Zh.\ Eksp.\ Teor.\ Fiz.\  {\bf 53}, 1717 (1967)].



\bibitem{Tanabashi:2018oca} 
  M.~Tanabashi {\it et al.} [Particle Data Group],
  Phys.\ Rev.\ D {\bf 98}, no. 3, 030001 (2018).



\bibitem{Esteban:2018azc} 
  I.~Esteban, M.~C.~Gonzalez-Garcia, A.~Hernandez-Cabezudo, M.~Maltoni and T.~Schwetz,
  JHEP {\bf 1901}, 106 (2019)
  [arXiv:1811.05487 [hep-ph]],
  NuFIT 4.0 (2018), www.nu-fit.org.



\bibitem{Zaglauer:1988gz}
  H.~W.~Zaglauer and K.~H.~Schwarzer,
  Z.\ Phys.\ C {\bf 40} (1988) 273.

\bibitem{Cervera:2000kp} 
  A.~Cervera, A.~Donini, M.~B.~Gavela, J.~J.~Gomez Cadenas, P.~Hernandez, O.~Mena and S.~Rigolin,
  Nucl.\ Phys.\ B {\bf 579}, 17 (2000)
  Erratum: [Nucl.\ Phys.\ B {\bf 593}, 731 (2001)]
  [hep-ph/0002108].

\bibitem{Freund:2001pn} 
  M.~Freund,
  Phys.\ Rev.\ D {\bf 64}, 053003 (2001)
  [hep-ph/0103300].

\bibitem{Xu:2015kma} 
  X.~J.~Xu,
  JHEP {\bf 1510}, 090 (2015)
  [arXiv:1502.02503 [hep-ph]].

\bibitem{Li:2016pzm} 
  Y.~F.~Li, J.~Zhang, S.~Zhou and J.~Y.~Zhu,
  JHEP {\bf 1612}, 109 (2016)
  [arXiv:1610.04133 [hep-ph]].

\bibitem{Denton:2018hal} 
  P.~B.~Denton and S.~J.~Parke,
  JHEP {\bf 1806}, 109 (2018)
  [arXiv:1801.06514 [hep-ph]].

\bibitem{Akhmedov:2004ny} 
  E.~K.~Akhmedov, R.~Johansson, M.~Lindner, T.~Ohlsson and T.~Schwetz,
  JHEP {\bf 0404}, 078 (2004)
  [hep-ph/0402175].



\bibitem{Barger:1980tf} 
  V.~D.~Barger, K.~Whisnant, S.~Pakvasa and R.~J.~N.~Phillips,
  Phys.\ Rev.\ D {\bf 22}, 2718 (1980).

\bibitem{Blennow:2003xw} 
  M.~Blennow, T.~Ohlsson and H.~Snellman,
  Phys.\ Rev.\ D {\bf 69}, 073006 (2004)
  [hep-ph/0311098].

\bibitem{Parke:2016joa} 
  S.~Parke,
  Phys.\ Rev.\ D {\bf 93}, no. 5, 053008 (2016)
  [arXiv:1601.07464 [hep-ph]].

\bibitem{Xing:2016ymg} 
  Z.~Z.~Xing and J.~Y.~Zhu,
  JHEP {\bf 1607}, 011 (2016)
  [arXiv:1603.02002 [hep-ph]].

\bibitem{Denton:2016wmg} 
  P.~B.~Denton, H.~Minakata and S.~J.~Parke,
  JHEP {\bf 1606}, 051 (2016)
  [arXiv:1604.08167 [hep-ph]].



\bibitem{Xing:2018lob} 
  Z.~Z.~Xing, S.~Zhou and Y.~L.~Zhou,
  JHEP {\bf 1805}, 015 (2018)
  [arXiv:1802.00990 [hep-ph]].

\bibitem{Huang:2018ufu} 
  G.~Y.~Huang, J.~H.~Liu and S.~Zhou,
  Nucl.\ Phys.\ B {\bf 931}, 324 (2018)
  [arXiv:1803.02037 [hep-ph]].

\bibitem{Wang:2019yfp} 
  X.~Wang and S.~Zhou,
  JHEP {\bf 1905}, 035 (2019)
  [arXiv:1901.10882 [hep-ph]].

\bibitem{Xing:2019owb} 
  Z.~Z.~Xing and J.~Y.~Zhu,
  Nucl.\ Phys.\ B {\bf 949}, 114803 (2019)
  [arXiv:1905.08644 [hep-ph]].





\bibitem{Xing:2000gg} 
  Z.~Z.~Xing,
  Phys.\ Lett.\ B {\bf 487}, 327 (2000)
  [hep-ph/0002246].

\bibitem{Xing:2001yg} 
  Z.~Z.~Xing,
  Phys.\ Rev.\ D {\bf 64}, 073014 (2001)
  [hep-ph/0107123].

\bibitem{Xing:2003ez}
  Z.~Z.~Xing,
  Int.\ J.\ Mod.\ Phys.\ A {\bf 19} (2004) 1
  [hep-ph/0307359].

\bibitem{Chiu:2010da} 
  S.~H.~Chiu, T.~K.~Kuo and L.~X.~Liu,
  Phys.\ Lett.\ B {\bf 687}, 184 (2010)
  [arXiv:1001.1469 [hep-ph]].

\bibitem{Zhou:2016luk} 
  S.~Zhou,
  J.\ Phys.\ G {\bf 44}, no. 4, 044006 (2017)
  [arXiv:1612.03537 [hep-ph]].

\bibitem{Chiu:2017ckv} 
  S.~H.~Chiu and T.~K.~Kuo,
  Phys.\ Rev.\ D {\bf 97}, no. 5, 055026 (2018)
  [arXiv:1712.08487 [hep-ph]].

















\bibitem{Minakata:2015gra} 
  H.~Minakata and S.~J.~Parke,
  JHEP {\bf 1601}, 180 (2016)
  [arXiv:1505.01826 [hep-ph]].




\bibitem{Parke:2018brr} 
  S.~J.~Parke, P.~B.~Denton and H.~Minakata,
  PoS NuFact {\bf 2017}, 055 (2018)
  [arXiv:1801.00752 [hep-ph]].

\bibitem{Denton:2018fex} 
  P.~B.~Denton, S.~J.~Parke and X.~Zhang,
  Phys.\ Rev.\ D {\bf 98}, no. 3, 033001 (2018)
  [arXiv:1806.01277 [hep-ph]].











\bibitem{Xing:2015fdg}
  Z.~Z.~Xing and Z.~H.~Zhao,
  Rept.\ Prog.\ Phys.\  {\bf 79} (2016) no.7,  076201
  [arXiv:1512.04207 [hep-ph]].



\bibitem{Jarlskog:1985ht} 
  C.~Jarlskog,
  Phys.\ Rev.\ Lett.\  {\bf 55}, 1039 (1985).

\bibitem{Wu:1985ea} 
  D.~d.~Wu,
  Phys.\ Rev.\ D {\bf 33}, 860 (1986).

\bibitem{Naumov:1991ju} 
V.~A.~Naumov,
Int.\ J.\ Mod.\ Phys.\ D {\bf 1}, 379 (1992).

\bibitem{Toshev:1991ku} 
  S.~Toshev,
  Mod.\ Phys.\ Lett.\ A {\bf 6}, 455 (1991).

\bibitem{Harrison:1999df} 
  P.~F.~Harrison and W.~G.~Scott,
  Phys.\ Lett.\ B {\bf 476}, 349 (2000)
  [hep-ph/9912435].

\bibitem{Yokomakura:2000sv} 
  H.~Yokomakura, K.~Kimura and A.~Takamura,
  Phys.\ Lett.\ B {\bf 496}, 175 (2000)
  [hep-ph/0009141].

\bibitem{Xing:2000ik} 
  Z.~z.~Xing,
  Phys.\ Rev.\ D {\bf 63}, 073012 (2001)
  [hep-ph/0009294].

\bibitem{Parke:2000hu} 
  S.~J.~Parke and T.~J.~Weiler,
  Phys.\ Lett.\ B {\bf 501}, 106 (2001)
  [hep-ph/0011247].


\bibitem{Denton:2019yiw} 
  P.~B.~Denton and S.~J.~Parke,
  Phys.\ Rev.\ D {\bf 100}, no. 5, 053004 (2019)
  [arXiv:1902.07185 [hep-ph]].
 
\bibitem{Wang:2019dal} 
  X.~Wang and S.~Zhou,
  Nucl.\ Phys.\ B {\bf 950}, 114867 (2020)
  [arXiv:1908.07304 [hep-ph]].







\bibitem{PREM}
A.~D.~Dziewonski and D.~L.~Anderson, Physics of the Earth and Planetary Interiors {\bf 25}, 297 (1981).

\bibitem{Stacey}
F.~D.~Stacey, {\it Physics of the Earth, 2nd edition}, John Wiley and Sons, London, New York, 1977.



%
%

\bibitem{Maris:1997nk} 
  M.~Maris and S.~T.~Petcov,
  Phys.\ Rev.\ D {\bf 56}, 7444 (1997)
  [hep-ph/9705392].



\bibitem{Gandhi:1995tf} 
  R.~Gandhi, C.~Quigg, M.~H.~Reno and I.~Sarcevic,
  Astropart.\ Phys.\  {\bf 5}, 81 (1996)
  [hep-ph/9512364].

\bibitem{Gandhi:1998ri} 
  R.~Gandhi, C.~Quigg, M.~H.~Reno and I.~Sarcevic,
  Phys.\ Rev.\ D {\bf 58}, 093009 (1998)
  [hep-ph/9807264].



\bibitem{Naumov:1998sf} 
  V.~A.~Naumov and L.~Perrone,
  Astropart.\ Phys.\  {\bf 10}, 239 (1999)
  [hep-ph/9804301].

\bibitem{Palomares-Ruiz:2015mka} 
  S.~Palomares-Ruiz, A.~C.~Vincent and O.~Mena,
  Phys.\ Rev.\ D {\bf 91}, no. 10, 103008 (2015)
  [arXiv:1502.02649 [astro-ph.HE]].

\bibitem{Vincent:2017svp} 
  A.~C.~Vincent, C.~A.~Argüelles and A.~Kheirandish,
  JCAP {\bf 1711}, 012 (2017)
  [JCAP {\bf 1711}, 012 (2017)]
  [arXiv:1706.09895 [hep-ph]].

\bibitem{Donini:2018tsg} 
  A.~Donini, S.~Palomares-Ruiz and J.~Salvado,
  Nature Phys.\  {\bf 15}, no. 1, 37 (2019)
  [arXiv:1803.05901 [hep-ph]].



\bibitem{Shapiro:1983du} 
  S.~L.~Shapiro and S.~A.~Teukolsky,
  ``Black holes, white dwarfs, and neutron stars: The physics of compact objects,''
  New York, USA: Wiley (1983) 645 p.

\bibitem{MaxCamenzind}
M. Camenzind, 
``Compact Objects in Astrophysics: White Dwarfs, Neutron Stars and Black Holes,''
In: 
Astronomy and Astrophysics Library, Springer-Verlag Berlin Heidelberg, Leipzig, 2007, pp. 355-572.


\bibitem{An:2015jdp} 
  F.~An {\it et al.} [JUNO Collaboration],
  J.\ Phys.\ G {\bf 43}, no. 3, 030401 (2016)
  [arXiv:1507.05613 [physics.ins-det]].
 
 
\bibitem{Giunti:2007ry} 
  C.~Giunti and C.~W.~Kim,
  Oxford, UK: Univ. Pr. (2007) 710 p.

\bibitem{Xing:2011zza} 
  Z.~z.~Xing and S.~Zhou,
  Springer-Verlag, Berlin Heidelberg (2011).



\end{thebibliography}
\end{document}